\documentclass[11pt,twoside,a4paper]{amsart}
\usepackage[totalwidth=385pt,totalheight=584pt]{geometry}
\usepackage{amsmath,amsthm,amssymb,epsf,graphics,graphicx,mathrsfs,yfonts,color}
\usepackage{psfrag}
\usepackage{amssymb}
\usepackage{tikz}
\usetikzlibrary{chains}
\newcommand{\bz}{\overline{z}}

\newtheorem{cj}{Conjecture}

\newcommand\blank[1]{}
\renewcommand{\hat}{\widehat}
\newcommand\eq{\begin{equation}}
\newcommand\en{\end{equation}}
\newcommand\bea{\begin{eqnarray}}
\newcommand\eea{\end{eqnarray}}

\newcommand\ba{\(\begin{array}}
\newcommand\ea{\end{array}\)}

\newcommand\bbeta{\boldsymbol{\beta}}
\newcommand\bgamma{\boldsymbol{\gamma}}

\newcommand\bmu{\boldsymbol{\mu}}

\begin{document}

\begin{title}{On sinh-Gordon Thermodynamic Bethe Ansatz and fermionic basis.}
\end{title}
\author{S.~Negro}

\address{SN
Department of mathematical sciences, University of Durham,
 Science Laboratories, South Rd, Durham DH1 3LE, United Kingdom. \hspace{1cm}
 Dip. di Fisica and INFN, Universit\`a di Torino, Via P. Giuria 1, 10125
Torino, Italy}\email{stefano.negro@durham.ac.uk\hspace{1cm} negro@to.infn.it}

\begin{abstract}
We review the construction of the fermionic basis for sinh-Gordon model and investigate numerically the ultra-violet limit of the one-point functions. We then compare the predictions obtained from this formalism against previously established results.
 \end{abstract}

\maketitle

\section{Introduction}	
\label{intro}

In the study of a Quantum Field Theory (QFT), the one-point functions play a fundamental {\it{r\^ole}}; indeed, when using the Operator Product Expansion (OPE) to calculate the ultraviolet asymptotics of a correlation function, one needs to know both the coefficients of the said expansion and the one-point functions of the local operators in the theory. While the former are purely ultra-violet objects and can, in principle, be extracted via perturbation theory of the corresponding ultra-violet Conformal Field Theory (CFT), the one-point functions depend essentially on the infra-red structure of the theory, where perturbative techniques are of no help at all. Thus the development of new methods to explore the infra-red region is of primary importance.

Integrable models are the perfect playground where one can experiment with new analytical methods aimed at extracting data; in particular the sinh-Gordon model is the simplest example of massive integrable QFT and, at the same time, is complicated enough to display interesting structures. Moreover this model, along with its twin, the sine-Gordon model, has received plenty of attention in the last 30 years and nowadays most of its features are known.

Computing one-point functions in an integrable deformation of a CFT is anything but an easy task and we wish to explain the reasons for this fact clearly. Although in our deformed theory the conformal invariance is broken, the local fields retain a one-to-one correspondence with those of the original CFT, which are organized according to the corresponding Virasoro algebra in the usual way. This means that in the perturbed theory there exist fields $\Phi_a(z,\bar z) = e^{a\eta(z,\bar z)}$ which can be deemed as \emph{primary}, whose space of descendants can be identified with the tensor product of Verma modules $\mathcal V_a \otimes \bar{\mathcal V}_a$ of the unperturbed CFT. The operators acting in the space of states of the perturbed CFT can thus be interpreted as operators acting on the corresponding Verma modules and, consequently, one-point functions appear to be functionals on the tensor product $\mathcal V_a \otimes \bar{\mathcal V}_a$. However, we still have not taken in account the integrable structure of the model; in fact \emph{all one-point functions of descendants built out of integrals of motion identically vanish}. This means that the correct space on which the one-point function should be defined as a linear functional is the tensor product $\mathcal V^{\textrm{quo}}_a \otimes \bar{\mathcal V}^{\textrm{quo}}_a$ of the two quotient spaces

\begin{equation}
	\mathcal V^{\textrm{quo}}_a \doteq \mathcal V_a \Bigg/\sum_{k=1}^\infty \mathbf i_{2k-1}\mathcal V_a \ , \qquad \bar{\mathcal V}^{\textrm{quo}}_a \doteq \bar{\mathcal V}_a \Bigg/\sum_{k=1}^\infty \bar{\mathbf i}_{2k-1} \bar{\mathcal V}_a \; ,
\end{equation}
where with $\mathbf i_{2k-1}$ (respectively $\bar{\mathbf{i}}_{2k-1}$) we denote the action of the chiral (antichiral) integrals of motion on the Verma module.

It's now becoming clear what is the main issue: the basis we introduced above, composed of the primary fields $\Phi_a(z,\bar z)$ and their ``conformal" descendants, is a basis for the full Verma module! In order to reduce this last to a basis of the quotient space, one has to factor out by hand all the null vectors which arise from the action of the integrals of motion and their form quickly becomes rather involved. One would rather work directly in the quotient space, where the factoring of null vectors is automatically taken in account, and fix uniquely a basis by means of some physical requirement. A basis of this kind was actually discovered some years ago for the six-vertex model \cite{Boos_Jimb_Miwa_Smir_Take_07,Boos_Jimb_Miwa_Smir_Take_09,Jimb_Miwa_Smir_09} and immediately extended to CFT \cite{Boos_Jimb_Miwa_Smir_10}, sine-Gordon \cite{Jimb_Miwa_Smir_10,Jimb_Miwa_Smir_11-1,Jimb_Miwa_Smir_11-2} and sinh-Gordon models \cite{Negr_Smir_13-2}.

The building blocks of this basis are the primary fields $\Phi_a(z,\bar z)$ and creation operators which, acting on the former, produce the descendants, much like what happens for the usual conformal basis; the peculiar fact is that these creation operators are \emph{fermions}. There are two of them for each chirality : $\bbeta^\ast_{2j-1}$, $\bgamma^\ast_{2j-1}$, $\bar{\bbeta}^\ast_{2j-1}$ and $\bar{\bgamma}^\ast_{2j-1}$. In the above-cited articles, these fermions were defined, in a mathematically rigorous fashion for six-vertex, CFT and sine-Gordon models and as an educated conjecture for the sinh-Gordon model, and their properties were thoroughly analysed; in particular for sin(h)-Gordon model\footnote{Here and in the following, the shorthand sin(h)-Gordon is used to denote both sine-Gordon and sinh-Gordon.}, the quotient space $\mathcal V^{\textrm{quo}}_a \otimes \bar{\mathcal V}^{\textrm{quo}}_a$ was shown to allow the following basis:

\begin{equation}
	\bbeta^{\ast}_{I^+} \bar\bbeta^{\ast}_{\bar I^+} \bar\bgamma^{\ast}_{\bar I^-} \bgamma^{\ast}_{I^-} \Phi_a(0) \ , \qquad \mathfrak C(I^+)=\mathfrak C(I^-) \ , \  \mathfrak C(\bar I^+)=\mathfrak C(\bar I^-) \; ,
\end{equation}
where $I^{\pm} = \{2i_1^\pm-1, \ldots , 2i_n^\pm-1\}$ and similarly for $\bar I^\pm$. The symbol $\mathfrak C(I)$ stands for the \emph{cardinality} of the set $I$ and the following multi-index notation is introduced:

\begin{equation}
	A_I = A_{i_1}A_{i_2}\ldots A_{i_n} \ ; \quad \vert I\vert \doteq \sum_{p=1}^{\mathfrak C(I)} i_p \ ; \quad aI+b = \{ai_1+b,\ldots,ai_n+b\} \; ,
\end{equation}
where $a,b \in \mathbb Z$ and $I=\{i_1,\ldots,i_n\}$.

While the rigorous construction of this basis, presented in Refs. \cite{Boos_Jimb_Miwa_Smir_10}-\cite{Jimb_Miwa_Smir_11-1}, might appear somewhat cumbersome and hard to understand, when the dust raised by their construction has fallen, the fermions reveal their true strength in the simple and beautiful determinant formula for the one-point functions:

\begin{equation}
	\frac{\langle\bbeta^{\ast}_{I^+} \bar\bbeta^{\ast}_{\bar I^+} \bar\bgamma^{\ast}_{\bar I^-} \bgamma^{\ast}_{I^-} \Phi_a(0)\rangle_R}{\langle\Phi_a(0)\rangle_R} = \mathcal D\Big(I^+\cup (-\bar I^+) \Big\vert I^-\cup (-\bar I^-) \Big\vert \alpha\Big) \; ,
\label{eq:detformula}
\end{equation}
where, for two sets $A=\{a_j\}_{j=1}^n$ and $B=\{b_j\}_{j=1}^n$, the function $\mathcal D$ is defined as follows

\begin{align}
	\mathcal D(A \vert B \vert \alpha ) &\doteq \left(\prod_{\ell=1}^n \frac{\textrm{sgn}(a_\ell)\textrm{sgn}(b_\ell)}{\pi}\right)\times	\nonumber
	\\
	 &\times\det \Big[ \Theta(i a_j, i b_k \vert\alpha ) - \pi\textrm{sgn}(a_j)t_{a_j}(\alpha) \; \delta_{a_j,-b_k} \Big]_{j,k=1}^n
\end{align}
and the functions $\Theta(a,b\vert \alpha)$ and $t_{a}(\alpha)$ will be defined below. The parameter $\alpha$ is related to the conformal dimension of the primary field by 

\begin{equation}
	\alpha = \frac{2}{b + b^{-1}} a \; .
\end{equation}

A very important property of the fermions is that, aside from allowing the construction of the descendants, they can be used in order to \emph{shift} the primary and descendant fields in their conformal dimension $a$. As it is shown in Ref. \cite{Jimb_Miwa_Smir_11-1}, if we give up the conditions $\mathfrak C(I^+)=\mathfrak C(I^-) , \  \mathfrak C(\bar I^+)=\mathfrak C(\bar I^-)$ in favour of the less restraining $\mathfrak C(I^+) - \mathfrak C(I^-) = \mathfrak C(\bar I^-) - \mathfrak C(\bar I^+) = m$, then the following relation holds

\begin{align}
	&\bbeta^\ast_{I^+}\bar\bbeta^\ast_{\bar I^+}\bar\bgamma^\ast_{\bar I^-}\bgamma^\ast_{I^-} \Phi_{a-mb}(0) \cong \nonumber
	\\
	&\cong \frac{C_m(a)}{\prod_{j=1}^m t_{2j-1}(a)}\bbeta^\ast_{I^+ +2m}\bar\bbeta^\ast_{\bar I^+ -2m}\bar\bgamma^\ast_{\bar I^- +2m}\bgamma^\ast_{I^- -2m}\bbeta^\ast_{I_{\textrm{odd}}(m)}\bar\bgamma^\ast_{I_{\textrm{odd}}(m)}\Phi_a(0)	\label{eq:shiftformula}
\end{align}
where $I_{\textrm{odd}}(m) = \{1,3,\ldots,2m-1\}$ and we use the symbol $\cong$ to denote identification in weak sense (that is, under expectation value).\\

As was mentioned above, for the sine-Gordon model the fermionic basis can be build in a mathematically rigorous fashion; the authors of Ref. \cite{Jimb_Miwa_Smir_11-1} performed this task by relying on the fact that sine-Gordon model allows for a lattice regularization in the form of the eight-vertex model, which is well studied and relatively easy to manage. Conversely, for its twin, the sinh-Gordon model, the situation is not so simple: the lattice regularization, in this case, takes the form of a much more complicated model, where the Boltzmann weights are defined in terms of the R-matrix of the tensor product of two infinite-dimensional representations of $U_q(\hat{\mathfrak{sl}}_2)$ \cite{Byts_Tesc_06,Tesc_07}. So far the status of the phase transition for this model has not been clarified and, thus, relying on the lattice regularization is not a viable strategy for sinh-Gordon model.

An alternative approach is to start directly from the Thermodynamic Bethe Ansatz (TBA) equations which, for the sinh-Gordon model, exhibit a very simple structure, given the fact that the spectrum of the theory consists of a single particle. This fact let the authors of Ref. \cite{Negr_Smir_13-2} straightforwardly define all the functions involved in the formula (\ref{eq:detformula}). However that same formula, along with the very existence of the fermionic basis, had to be introduced as a conjecture based on two facts:

\begin{itemize}
	\item 
	From purely algebraic point of view, the ultra-violet (UV) limit of the sinh-Gordon model corresponds to the CFT considered in Ref. \cite{Jimb_Miwa_Smir_09}; this last theory is, at the same time, the UV limit of the sine-Gordon model.
	\item There are two possible interpretations of the sin(h)-Gordon action, as a perturbation of the free boson CFT\footnote{Here and in the following the notation $\partial = \frac{\partial}{\partial z}$ and $\bar\partial = \frac{\partial}{\partial \bar z}$ will be used.}:
		\begin{equation}
			\mathcal A = \int \left\{\left[\frac{1}{4\pi}\partial\eta(z,\bar z) \bar\partial\eta(z,\bar z)\right] + \frac{2\bmu^2}{\sin(\pi b^2)}\cosh[b\, \eta(z,\bar z)]\right\}\frac{dz\wedge d\bar z}{2} \; ,	\nonumber
		\end{equation}
	or as a perturbation of the Liouville model, conventionally identified as the minimal CFT with central charge $c=1+6Q^2$, where $Q=b+b^{-1}$:
		\begin{equation}
			\mathcal A = \int \left\{\left[\frac{1}{4\pi}\partial\eta(z,\bar z) \bar\partial\eta(z,\bar z) + \frac{\bmu^2}{\sin(\pi b^2)} e^{b\, \eta(z,\bar z)}\right] + \frac{\bmu^2}{\sin(\pi b^2)}e^{-b\, \eta(z,\bar z)}\right\}\frac{dz\wedge d\bar z}{2} \; .	\nonumber
		\end{equation}
	This twofold interpretation of the action led the authors of Ref.  \cite{Fate_Frad_Luky_AZam_AlZa_99} to some functional relations for the one-point functions of sine-Gordon model, which were named \emph{reflection relations}. In Ref. \cite{Negr_Smir_13-1} it was shown how the fermionic basis can be interpreted as a basis of the space of states for which these reflection relations are trivially satisfied.
\end{itemize}
A remark about the choice for the normalization of the dimensional constant is necessary. As discussed in Ref. \cite{Negr_Smir_13-2}, this choice, aside from being extremely convenient for the calculations, encloses serious physical reasons. Firstly it takes automatically into account the change of sign in the potential energy when passing from sinh- to sine-Gordon and encodes also the pole at $b = i$ of this last\footnote{Due to the fact that the perturbing operator becomes irrelevant for $b^2 < -1$. Note that there are poles also for $b \in \mathbb Z$ which look natural once one consider the physical scale of the model, namely the mass of the particle \cite{AlZa_06}.}; more importantly,
this normalisation allows the mass m of both the sinh-Gordon particle and that of
the sine-Gordon lowest breather to be expressed by a universal formula:

\begin{equation}
	\bmu\Gamma(1+b^2) = \left[\frac{m}{4\sqrt{\pi}} \Gamma\left(\frac{1}{2(1+b^2)}\right) \Gamma\left(1+\frac{b^2}{2(1+b^2)}\right)\right]^{1+b^2} \; .
\label{eq:massformula}
\end{equation}

\vspace{1cm}
Since, for the sinh-Gordon model, the formula (\ref{eq:detformula}) and the existence of the fermionic basis still retain the status of conjectures, it is of utmost importance to obtain {\it a posteriori} confirmations of their validity, by checking the predictions against known results. Analytic comparison with results of Refs. \cite{Lecl_Muss_99} and \cite{Luky_01} were already performed in Ref. \cite{Negr_Smir_13-2}.

The purpose of this paper is to obtain further confirmations of the validity of (\ref{eq:detformula}), by means of numerical simulations. In particular the one-point functions of the sinh-Gordon model, defined on a cylinder of radius $2\pi R$, were numerically evaluated for very small values of the radius $R \sim 0$, limit in which the model approaches its UV limit; these numerical results were then compared against the theoretical behaviours obtained in Refs.  \cite{Boos_Jimb_Miwa_Smir_10} and \cite{Luky_01}. As in the last-cited article a rescaling of the model to a circumference of fixed radius $2 \pi$ is to be performed; this amounts to a renormalisation of the physical mass $m\rightarrow m R$, so that $\bmu \propto R^{1+b^2}$.

It has to be noted that, since the goal is to compare the results obtained from (\ref{eq:detformula}) with the known ``CFT behaviour", it is wise to avoid the possible complications arising in the regions of the parameter space where $a-b<0$. Let us clarify this point.

Looking at the formula for the conformal dimension shift of the fields (\ref{eq:shiftformula}) we see that the ratio of expectation values of the two primary fields $\Phi_{a-b}(0)$ and $\Phi_a(0)$ can be expressed in terms of the ratio of the one-point function of the descendant $\bbeta_1^\ast\bar\bgamma_1^\ast\Phi_a(0)$ with that of the primary field $\Phi_a(0)$; in formulae

\begin{equation}
	\frac{\langle\Phi_{a-b}(0)\rangle}{\langle\Phi_{a}(0)\rangle} = \frac{C_1(a)}{t_1(a)} \frac{\langle\bbeta_1^\ast\bar\bgamma_1^\ast\Phi_a(0)\rangle}{\langle\Phi_{a}(0)\rangle} \; .
\label{eq:primarytodescendant}
\end{equation}
As one approaches the UV limit $R\sim 0$, the one-point functions of primary fields are believed to behave as three-point functions of the Liouville CFT \cite{Luky_01} with two additional fields, of dimensions $\Delta_{\pm}=\frac{Q^2}{4}-P(R)^2$ where $P(R)$ is the quantized momentum of Liouville CFT, placed at $\pm\infty$. This, however, holds true only if the dimensions of the fields are positive, which means

\begin{equation}
	\left\{\begin{array}{l}
				0 < a < Q \\
				0 < a-b < Q
			\end{array}\right.
	\quad \Rightarrow \quad b < a < Q \; .
\end{equation}

This fact becomes evident, for example, sending $a\rightarrow 0$; in this case, the expectation value of the field $\Phi_{-b}(0) = e^{-b\eta(0)}$ can be calculated directly in terms of the ground-state energy $E(R) \underset{R\rightarrow 0}{\sim} -\frac{\pi}{6 R}c_{\textrm{eff}}(R)$ where \cite{Luky_01}

\begin{equation}
	c_{\textrm{eff}}(R) \underset{R\rightarrow 0}{\sim} 1 - \frac{24\pi}{\Big(\delta_1-4Q\log\frac{R}{2\pi}\Big)^2} \; ,
\end{equation}
and $\delta_1$ is a constant.
With (\ref{eq:detformula}), (\ref{eq:primarytodescendant}) and the relation between the function $\Theta(i,-i\vert 0)$ and $E(R)$ shown in Ref.  \cite{Jimb_Miwa_Smir_11-1}, one obtains:

\begin{equation}
	\langle e^{-b\eta(0)}\rangle = -\frac{C_1(0)}{\pi m^2 t_1(0)} \left(\frac{1}{R}+\frac{d}{dR}\right)E(R) \underset{R\rightarrow 0}{\sim} \frac{\pi c_1(0,b)}{2m^2Q^2} \frac{R^{-2(b^2+1)}}{(-\log R)^3} \; ,
\end{equation}
where the definition (\ref{eq:C-constants}) was applied, setting $C_1(a)/t_1(a)\underset{R\rightarrow 0}{\sim} c_1(a,b)R^{2b(2a-b)}$, $c_1(a,b)$ being a function of $a$ and $b$ only.
On the other hand, using the formula for the Liouville three-point amplitude (\ref{eq:liouvillethreepoint}) found in Refs. \cite{Dorn_Otto_94} and \cite{AZam_AlZa_96}, the result is radically different

\begin{align}
	&\langle e^{-b\eta(0)}\rangle = \frac{\langle\Phi_{\frac{Q}{2}-P}(-\infty)\vert\Phi_{-b}(0)\vert\Phi_{\frac{Q}{2}+P}(\infty)\rangle}{\langle\Phi_{\frac{Q}{2}-P}(-\infty)\vert\Phi_{\frac{Q}{2}+P}(\infty)\rangle} \underset{R\rightarrow 0}{\sim} k(a,b) R^{2(1+b^2)} \; .
\end{align}

It is clear that outside the natural region $b<a<Q$, the sinh-Gordon model do no more approaches na\"\i vely the Liouville CFT: there are contributions not taken in account which become important. However, as said above, rather than exploring the UV limit of the sinh-Gordon model {\it per se}, the goal of this paper is to use it in order to obtain evidence of the agreement between the predictions obtained from the fermionic basis and the results known in the literature: for this reason from now on the parameter space will be restricted to the region $0<b<a<Q$.

\section{The fermionic basis}
\label{sec:fermbasis}

Let us review briefly the properties of the fermionic basis.\\The two-fold interpretation of the sinh-Gordon action that we mentioned above has an interesting and important consequence. If we look at sinh-Gordon as a deformation of the free boson CFT, then the natural choice for the descendants of the primary field $\Phi_a(0)= e^{a\eta(0)}$ are normal ordered products of $e^{a\eta(0)}$ with polynomials of even degree\footnote{We limit ourselves to even degree polynomials, since $\mathcal V_a^{\textrm{quo}}$ non-trivial subspaces are of even dimension only.} in the derivatives of $\eta(0)$. In this \emph{Heisenberg basis} the one-point functions inherit the natural free boson symmetry

\begin{equation}
	\sigma_1 \; : a \rightarrow -a \; .
\end{equation}
On the other hand, if we consider the sinh-Gordon model as a deformation of the Liouville CFT, the descendants of $\Phi_a(0)$ are more naturally defined as normal-ordered products of $e^{a\eta(0)}$ with polynomials in even-degree derivatives of $T(z,\bz)$ and $\bar T(z,\bz)$, where

\begin{align}
	T(z,\bz) \doteq T_{z,z}(z,\bz)= -\frac{1}{4}\Big[\partial \eta(z,\bz)\Big]^2 + \frac{Q}{2}\partial^2\eta(z,\bz) \: ,	\nonumber
	\\
	\\
	\bar T(z,\bz) \doteq T_{\bz,\bz}(z,\bz)= -\frac{1}{4}\Big[\bar \partial \eta(z,\bz)\Big]^2 + \frac{Q}{2}\bar \partial^2\eta(z,\bz) \: ,	\nonumber
\end{align}
are the components of Liouville energy-momentum tensor. It is natural to assume that in this basis, that we call the \emph{Virasoro basis}, the one-point functions retain the symmetry of the Liouville model

\begin{equation}
	\sigma_2 \; : a \rightarrow Q-a \; .
\end{equation}

Since both the Heisenberg and the Virasoro basis are, when the action of the integrals of motion has been factored out, fully fledged bases of sinh-Gordon space of states, the one-point functions in any possible basis have to transform in some definite way under the symmetries $\sigma_1$ and $\sigma_2$. This fact gives rise to the above-mentioned reflection relations and suggests that there must exist a basis in which both these symmetries act in a simple, multiplicative way: this particular basis is the \emph{fermionic basis}; we consider then the fermions as defined by their behaviour under the symmetries $\sigma_1$ and $\sigma_2$. Starting from the Liouville CFT, where the fermions $\bbeta^{\textrm{CFT}\,\ast}$ and $\bgamma^{\textrm{CFT}\,\ast}$ can be defined as an intrinsic property of the model \cite{Negr_Smir_13-2}, we see that the reflections act on the fermionic basis as follows:

\begin{align}
  &\phantom{\sigma_1 \; : \; } \bgamma^{\textrm{CFT}\,\ast}_{2m-1} \rightarrow u(a)\bbeta^{\textrm{CFT}\,\ast}_{2m-1} 
  \phantom{\qquad , \qquad \sigma_2 \; : \qquad\; } \bgamma^{\textrm{CFT}\,\ast}_{2m-1} \rightarrow \bbeta^{\textrm{CFT}\,\ast}_{2m-1}	\nonumber
  \\
  &\sigma_1 \; : \;  \phantom{\bbeta^{\textrm{CFT}\,\ast}_{2m-1} \rightarrow u^{-1}(-a)\bbeta^{\textrm{CFT}\,\ast}_{2m-1}} \qquad , \qquad
  \sigma_2 \; : \;  
  \\
  &\phantom{\sigma_1 \; : \; } \bbeta^{\textrm{CFT}\,\ast}_{2m-1} \rightarrow u^{-1}(-a)\bgamma^{\textrm{CFT}\,\ast}_{2m-1}
  \phantom{\qquad , \qquad \sigma_2 \; : \;\; }\bbeta^{\textrm{CFT}\,\ast}_{2m-1} \rightarrow \bgamma^{\textrm{CFT}\,\ast}_{2m-1}	\nonumber
\end{align}
where

\begin{equation}
  u(a) \doteq \frac{-2a+b(2m-1)}{2a+b^{-1}(2m-1)} = \frac{-Q \alpha +b(2m-1)}{Q \alpha + b^{-1}(2m-1)}
\end{equation}
and for the second chirality we only have to change $a$ in $-a$ in the above function. There is an additional symmetry which was considered in Ref.
\cite{Negr_Smir_13-2}, that is the duality $b \rightarrow b^{-1}$, under which our fermions simply exchange

\begin{align}
  &\phantom{\textrm{duality} \quad\; } \bbeta^{\textrm{CFT}\,\ast}_{2m-1} \rightarrow \bgamma^{\textrm{CFT}\,\ast}_{2m-1}	\nonumber
  \\
  &\textrm{duality} \; :	\label{eq:duality}
  \\
  &\phantom{\textrm{duality} \quad\; } \bgamma^{\textrm{CFT}\,\ast}_{2m-1} \rightarrow \bbeta^{\textrm{CFT}\,\ast}_{2m-1}	\nonumber
\end{align}
The normalization of the fermions is such that when expressing the descendants in fermionic basis in terms of Virasoro descendants we have

\begin{equation}
  \bbeta^{\textrm{CFT}\,\ast}_{I^+}\bgamma^{\textrm{CFT}\,\ast}_{I^-}\Phi_a = C_{I^+,I^-} \left\{\mathbf{l}^n_{-2}+\cdots\right\}\Phi_a
  \; ,\qquad \mathfrak C(I^+) = \mathfrak C(I^-) = n \; ,
\end{equation}
with $\mathbf{l}_{n}$ being the coefficients of the Laurent expansion of the Liouville energy-momentum tensor component $T(z,\bz)$ while $C_{I^+,I^-}$ is the determinant of the Cauchy matrix $\{ 1/(i^+_j + i^-_k -1) \}_{j,k = 1}^n$.

The fermions for the sinh-Gordon model are obtained from the CFT ones simply by multiplication by a constant:

\begin{align}
  & \bbeta^{\ast}_{2m-1} = D_{2m-1}(a) \bbeta^{\textrm{CFT}\,\ast}_{2m-1} \qquad , \qquad 
  \bgamma^{\ast}_{2m-1} = D_{2m-1}(Q-a) \bgamma^{\textrm{CFT}\,\ast}_{2m-1} \; ,	\nonumber
  \\	\label{eq:cfttosinhfermions}
  \\ & \overline \bgamma^{\ast}_{2m-1} = D_{2m-1}(a) \overline\bgamma^{\textrm{CFT}\,\ast}_{2m-1} \qquad , \qquad 
  \overline \bbeta^{\ast}_{2m-1} = D_{2m-1}(Q-a) \overline\bbeta^{\textrm{CFT}\,\ast}_{2m-1} \; ,	\nonumber
\end{align}
where

\begin{equation}
  D_{2m-1}(a) = \frac{1}{2\pi i} \left(\frac{\bmu\Gamma(1+b^2)}{b^{1+b^2}}\right)^{-\frac{2m-1}{1+b^2}}\frac{\Gamma\left(\frac{a}{Q}+ \frac{2m-1}{2 b Q}\right)
  \Gamma\left(\frac{Q-a}{Q}+b\frac{2m-1}{2Q}\right)}{(m-1)!} \; .
\end{equation}
Note that this definition for the constants $D_{2m-1}$ differs from the one used in Refs. \cite{Boos_Jimb_Miwa_Smir_10}-\cite{Jimb_Miwa_Smir_11-1} by the factor

\begin{equation}
(-1)^m\sqrt{\frac{1+b^2}{i}}\frac{\bmu^{-\frac{2m-1}{1+b^2}}}{2\sin\left[\pi\left(\frac{a}{Q}-b\frac{2m-1}{2Q}\right)\right]} \; .
\end{equation}
The reason for this choice is twofold: on one side, the presence of $\bmu^{-\frac{2m-1}{1+b^2}}$ makes the fermions dimensionless while, on the other, the Q-periodic $\sin\left[\pi\left(\frac{a}{Q}-b\frac{2m-1}{2Q}\right)\right]$ lets the non-CFT fermions inherit the duality (\ref{eq:duality}). Of course this last holds iff the following term is ``self-dual''

\begin{equation}
  \frac{[\bmu\Gamma(1+b^2)]^{\frac{1}{1+b^2}}}{b}\; ,
\end{equation}
but this follows automatically when expressing $\bmu$ in terms of the sinh-Gordon particle mass, which is explicitly self-dual:

\begin{equation}
  \bmu \Gamma(1+b^2) = \left[ \frac{m}{4\sqrt{\pi}} \Gamma\left(\frac{1}{2(1+b^2)}\right)\Gamma\left(1+\frac{b^2}{2(1+b^2)}\right) \right]^{1+b^2} \; .
\end{equation}

The constants $t_{\ell}(a)$ and $C_m(a)$ introduced in (\ref{eq:shiftformula}) are defined as follows

\begin{equation}
  t_{\ell}(a) \doteq -\frac{1}{2}\sin^{-1}\left[\frac{\pi}{Q}\left(	 2a +\frac{\ell}{b}\right)\right]
\end{equation}

\begin{align}
  &C_m(a) \doteq \prod_{j=0}^{m-1} C_1(a-2bj) \; ,	\nonumber
  \\ \label{eq:C-constants}
  \\
  &C_1(a) \doteq [\bmu\Gamma(1+b^2)]^{4 x} \frac{\gamma(x)\gamma\left(\frac{1}{2}-x\right)}{2bQ\gamma(2 b x Q)} \; ,	\nonumber
\end{align}
where $2 Q x = 2 a -b$ and we denote $\gamma(y) \doteq \Gamma(y)/\Gamma(1-y) \; , \ \forall y \in \mathbb C$.

\section{TBA and one-point functions}
\label{sec:tbaandonepoint}

As said above, since the TBA equations for the sinh-Gordon model are extremely simple, it is quite straightforward to chose them as a starting point and proceed to the construction of the function $\Theta(l,m\vert\alpha)$ relying on the consistency equations which derive from the symmetries of the fermions.
Let us consider the sinh-Gordon model defined on an infinite cylinder of circumference $2\pi R$; we call the infinite direction the \emph{space direction} and the compact one the \emph{Matsubara direction}. The TBA for this model consists of a single integral equation:

\begin{equation}
  \epsilon(\theta) = 2 \pi m R\cosh\theta - \int\limits_{-\infty}^\infty \Phi(\theta-\theta') \log\left(1+ e^{-\epsilon(\theta')}\right) d\theta' \; ,
  \label{eq:destridevegaequation}
\end{equation}
with

\begin{align}
  \Phi(\theta) &= \frac{1}{2\pi\cosh\left(\theta + \pi i \frac{b^2-1}{2(b^2+1)}\right)} + \frac{1}{2\pi\cosh\left(\theta - \pi i \frac{b^2-1}{2(b^2+1)}\right)} = \int\limits_{-\infty}^\infty e^{ik\theta}\hat{\Phi}(k) \frac{dk}{2\pi} \; ,	\nonumber
  \\
  \\
  \hat{\Phi}(k) &= \frac{\cosh\left(\pi\frac{b^2-1}{2(b^2+1)}k\right)}{\cosh\left(\pi\frac{k}{2}\right)} \; .	\nonumber
\end{align}

Starting from this basic equation, one can build Baxter $Q$-functions in the Matsubara direction; namely define

\begin{equation}
  \log Q(\theta) = -\frac{\pi m R \cosh\theta}{\sin\frac{\pi}{b^2+1}} + \int\limits_{-\infty}^{\infty} \frac{\log\left(1+e^{-\epsilon(\theta')}\right)}{\cosh(\theta-\theta')}\frac{d\theta'}{2\pi} \; ,
\end{equation}
where we have chosen the first term on the right-hand side for consistency. It is straightforward to check that

\begin{equation}
  e^{-\epsilon(\theta)} = Q\left(\theta + \pi i\frac{b^2-1}{2(b^2+1)}\right) Q\left(\theta - \pi i\frac{b^2-1}{2(b^2+1)}\right) \; ,
\end{equation}
from which, recalling the Dirac delta representation $\cosh(\theta+i\frac{\pi}{2})+\cosh(\theta-i\frac{\pi}{2}) = 2\pi \delta(\theta)$, one can derive the bilinear equation\footnote{Actually, one should be careful and define correctly the analyticity conditions for the function $Q(\theta)$; a discussion can be found in Ref. \cite{AlZa_00}.}

\begin{equation}
  Q\left(\theta + \frac{\pi i}{2}\right)Q\left(\theta - \frac{\pi i}{2}\right) -Q\left(\theta + \pi i\frac{b^2-1}{2(b^2+1)}\right) Q\left(\theta - \pi i\frac{b^2-1}{2(b^2+1)}\right) =1 \; .
  \label{eq:quantumwronskian}
\end{equation}

Introducing $\zeta = e^{(b^2+1)\theta}$, it's not difficult to see how (\ref{eq:quantumwronskian}) implies that the function $T(\zeta)$, defined from the equation

\begin{equation}
  T(\zeta)Q(\theta) = Q\left(\theta +\pi i \frac{b^2}{b^2+1}\right) + Q\left(\theta -\pi i \frac{b^2}{b^2+1}\right) \; ,
\end{equation}
is a single-valued function of $\zeta^2$, with essential singularities at $\zeta = 0$ and $\zeta = \infty$. This equation is a second order finite difference equation for the function $Q(\theta)$ and thus admits two different solutions: $Q(\theta)$ and $Q(\theta+i\frac{\pi}{b^2+1})$, the equation (\ref{eq:quantumwronskian}) being their quantum Wronskian.

It's important to stress that the equations for the functions $Q(\theta)$ and $T(\theta)$ given here are to be considered as \emph{definitions}, so one should check that they are reasonable. A verification of the correctness of these definition was carried out in Ref. \cite{Luky_01}, where the behaviour of $T(\zeta)$ in the ultraviolet region $R\rightarrow 0$ is investigated numerically, showing how the asymptotics of $T(\zeta)$ for $\zeta \rightarrow 0$ and for $\zeta \rightarrow \infty$ correcly reproduce the eigenvalues of CFT integrals of motion and, moreover, that their normalisation is the same as in the sine-Gordon case \cite{Bazh_Luky_AZam_96,Bazh_Luky_AZam_97,Bazh_Luky_AZam_99}; this is an extremely convincing argument.

Now, having the TBA equation (\ref{eq:destridevegaequation}) at our disposal, we introduce a deformed kernel $\Phi_{\alpha}(\theta)$ requiring that its Fourier image $\hat{\Phi}(k,\alpha)$ satisfy $\hat\Phi(k,0) = \hat\Phi(k)$ and the following symmetries

\begin{align}
  \hat{\Phi}(k,\alpha+&2) = \hat{\Phi}(k,\alpha) \; ,\qquad \hat{\Phi}(k,-\alpha) = \hat{\Phi}(-k,\alpha) \; , \nonumber
 \\
 \\ &\hat{\Phi}(k,\alpha - 2\frac{b^2}{b^2+1}) = \hat{\Phi}(k+2i,\alpha)\; .	\nonumber
\end{align}
The first two relations directly derive from the request that the fermions transform in the correct way under the transformations $\sigma_1$ and $\sigma_2$; the third one, on the other hand, is necessary in order to grant the validity of the shift relation (\ref{eq:shiftformula}), as was shown in Ref. \cite{Negr_Smir_13-2}.

It's not hard to find that the kernel we're looking for has the following form:

\begin{align}
  &\Phi_{\alpha}(\theta) = \frac{e^{i\pi\alpha}}{2\pi\cosh\left(\theta +\pi i\frac{b^2-1}{2(b^2+1)}\right)}+\frac{e^{-i\pi\alpha}}{2\pi\cosh\left(\theta -\pi i\frac{b^2-1}{2(b^2+1)}\right)} = \int\limits_{-\infty}^\infty e^{ik\theta}\hat{\Phi}(k,\alpha)\frac{dk}{2\pi} \; ,	\nonumber
  \\
  \\
  &\hat{\Phi}(k,\alpha) = \frac{\cosh\left(\pi\frac{b^2-1}{2(b^2+1)}k - i \pi \alpha\right)}{\cosh\left(\pi\frac{k}{2}\right)} \; .	\nonumber
\end{align}
It is interesting to notice that, contrary to the function $\hat R(k,\alpha)$ of the sine-Gordon model \cite{Jimb_Miwa_Smir_11-1}, the deformed kernel $\hat\Phi$ does not have poles in the $k$-plane whose positions depend on $\alpha$. This simplification in the kernel structure is directly correlated to the fact that sinh-Gordon one-point functions, as functions of $\alpha$, have much simpler analytical properties than those of sine-Gordon.

Let us proceed by defining the dressed resolvent, which satisfies to the equation 

\begin{equation}
  R_{\textrm{dress}}(\theta,\theta'\vert\alpha) - \left[\Phi\ast \, R_{\textrm{dress}}\right](\theta,\theta'\vert\alpha) = \Phi(\theta,\theta'\vert\alpha) \; ,
\end{equation}
where $\Phi(\theta,\theta'\vert\alpha) \equiv \Phi_\alpha(\theta-\theta')$ and the $\ast$ denotes a deformed convolution

\begin{equation}
  [f\ast \, g](\theta,\theta') \doteq \int\limits_{-\infty}^\infty f(\theta,\phi)g(\phi,\theta')dm(\phi) \; ,\qquad dm(\phi) \doteq \frac{d\phi}{1+e^{\epsilon(\phi)}} \; .
\end{equation}

Now, using the dressed resolvent, we build the function $\Theta^{\textrm{shG}}_R$:

\begin{equation}
  R_{\textrm{dress}}(\theta,\theta'\vert\alpha) - \Phi_\alpha(\theta-\theta') = \int\limits_{-\infty}^\infty\int\limits_{-\infty}^\infty \frac{dl}{2\pi}\frac{dm}{2\pi} \hat{\Phi}(l,\alpha)\Theta^{\textrm{shG}}_R(l,m\vert\alpha)\hat \Phi(m,-\alpha)e^{il\theta+im\theta'} \; .
\end{equation}
Straightforward calculations show that the function $\Theta^{\textrm{shG}}_R$ satisfies the following equation

\begin{equation}
  \Theta^{\textrm{shG}}_R(l,m\vert\alpha) - G(l+m) - \int\limits_{-\infty}^\infty G(l-k)\hat \Phi(k,\alpha) \Theta^{\textrm{shG}}_R(k,m\vert\alpha) \frac{dk}{2\pi} = 0 \; ,
\end{equation}
with the function $G(k)$ being the $k$-moment of the measure $dm(\theta)$

\begin{equation}
  G(k) \doteq \int\limits_{-\infty}^\infty e^{-ik\theta} \frac{d\theta}{1+e^{\epsilon(\theta)}} \; .
\end{equation}

A useful way to express the function $\Theta^{\textrm{shG}}_R$ is the following

\begin{equation}
  \Theta^{\textrm{shG}}_R(il,im\vert\alpha) = e_l\ast \, e_m + e_l\ast \, R_{\textrm{dress}}^{(\alpha)} \ast \, e_m \; ,
\label{eq:sinhthetaerepresentation}
\end{equation}
where we have introduced the shorthand notation $e_l(\theta) \doteq e^{l\,\theta}$.

Since, for the ground state, the function $\epsilon(\theta)$ is even, from the symmetries of $\hat{\Phi}(k,\alpha)$ one easily derives the following relations:

\begin{equation}
  \Theta^{\textrm{shG}}_R(l,m\vert - \alpha) = \Theta^{\textrm{shG}}_R(m,l\vert\alpha) \; ,\qquad \Theta^{\textrm{shG}}_R(l,m\vert\alpha +2) = \Theta^{\textrm{shG}}_R(l,m\vert\alpha) \; ,
\end{equation}

\begin{align}
  \Theta^{\textrm{shG}}_R(l,m\vert \alpha - 2\frac{b^2}{b^2+1}) &- \Theta^{\textrm{shG}}_R(l+2i,m-2i\vert\alpha) =	\nonumber
  \\
  & = \frac{\Theta^{\textrm{shG}}_R(l+2i,-i\vert\alpha)\Theta^{\textrm{shG}}_R(i,m-2i\vert\alpha)}{ \pi t_1(\frac{Q}{2}\alpha) - \Theta^{\textrm{shG}}_R(i,-i\vert\alpha)} \; .	 \label{eq:thetashift}
\end{align}

As has been said in the introduction, the function $\Theta_R^{\textrm{shG}}$ can be used in order to calculate the expectation values of descendants in the fermionic basis:

\begin{cj}
  We conjecture that, as in the sine-Gordon model, the one-point functions of the sinh-Gordon model in the fermionic basis are expressed in terms of a determinant
  
  \begin{equation}
    \frac{\langle\bbeta^\ast_{I^+}\overline\bbeta^\ast_{\overline I^+}\overline\bgamma^\ast_{\overline I^-}\bgamma^\ast_{I^-}\Phi_\alpha(0)\rangle_R}{\langle\Phi_\alpha(0)\rangle_R} = \mathcal D\left(I^+ \cup(-\overline I^+) \vert I^-\cup(-\overline I^-)\vert\alpha\right) \; ,
    \label{eq:onepointfunctionsinh}
  \end{equation}
  where, for two sets $A=\{a_j\}_{j=1}^n$ and $B=\{b_j\}_{j=1}^n$, we have

\begin{equation}
  \mathcal D(A\vert B\vert \alpha) \doteq \left(\prod_{\ell=1}^n \frac{\textrm{sgn}(a_\ell)\textrm{sgn}(b_\ell)}{\pi} \right) 
  \det \left[\Theta^{\textrm{shG}}_R(ia_j,ib_k\vert \alpha)-\pi \delta_{a_j,-b_k} \textrm{sgn}(a_j) t_{a_j}(\alpha)\right]_{j,k=1}^n
\end{equation}
\label{conj:conj}
\end{cj}
Notice how, since $\Theta^{\textrm{shG}}_R(l,m\vert\alpha)\underset{R\rightarrow\infty}{\rightarrow}0$, in the infinite volume limit $R\rightarrow\infty$ the formulae for the one-point functions in sinh-Gordon coincide with the analytic continuation with respect to $b$ of the corresponding ones in sine-Gordon model \cite{Jimb_Miwa_Smir_11-1}.

\section{Numerical analysis in the $R\rightarrow 0$ limit}
\label{sec:numan}

We now turn to the numerical evaluation of the one-point functions of the sinh-Gordon model in the UV limit $R\rightarrow 0$. We will begin by studying the behaviour of the descendant fields and then move to the primary ones. As mentioned in the introduction, we rescale the model on a cylinder of radius $2\pi$ and take $r=2\pi m R$ as the parameter to be sent to zero.

\subsection{Descendant fields}
\label{subsec:descendant}

We are interested in the UV behaviour of the following class of one-point functions

\begin{equation}
	F_{2j-1,2k-1}(\alpha,r) \doteq \frac{\langle\bbeta^\ast_{2j-1}\bgamma^\ast_{2k-1}\Phi_\alpha\rangle_r}{\langle\Phi_\alpha\rangle_r} \; , \qquad j,k \in \mathbb N \; ,
\end{equation}
which can be rewritten using (\ref{eq:cfttosinhfermions}) as

\begin{equation}
	F_{2j-1,2k-1}(\alpha,r) = D_{2j-1}(\alpha) D_{2k-1}(2-\alpha) \frac{\langle\bbeta^{\textrm{CFT}\,\ast}_{2j-1} \bgamma^{\textrm{CFT}\,\ast}_{2k-1} \Phi_\alpha(0)\rangle_r}{\langle\Phi_\alpha(0)\rangle_r} \; , \nonumber
\end{equation}
In the $r\rightarrow 0$ limit, these functions should behave like ratios of CFT one-point functions. In particular, using the formulae found in the appendix of Ref. \cite{Jimb_Miwa_Smir_11-1}, we see that

\begin{equation}
	F_{2j-1,2k-1} \underset{r\rightarrow 0}{\sim} -\left(\frac{2 \pi m}{r}\right)^{2j+2k-2} \frac{D_{2j-1}(\alpha)D_{2k-1}(2-\alpha)}{j+k-1}\Omega_{2j-1,2k-1} \; ,
\label{eq:theodesc}
\end{equation}
where $\Omega_{2j-1,2k-1}$ are functions of the vacuum eigenvalues $I_{2n-1}$ of the integrals of motion, which can be found, for example, in Ref. \cite{Bazh_Luky_AZam_97}. For the cases we are interested in we have

\begin{align}
	&\Omega_{1,1}(\alpha ,r) = I_1(r) - \frac{\Delta_\alpha}{12} \; , \nonumber
	\\
	\\
	&\Omega_{1,3}(\alpha ,r) = I_3(r) - \frac{\Delta_\alpha}{6} I_1(r) + \frac{\Delta_\alpha^2}{144} + \frac{c+5}{1080}\Delta_\alpha - \frac{\Delta_\alpha}{360} d_\alpha \; , \nonumber
\end{align}
where

\begin{equation}
	\Delta_\alpha = \frac{Q^2}{4}\alpha(2-\alpha) \; , \qquad d_\alpha = \frac{1}{6}\sqrt{(25 - c)(24\Delta_\alpha+1-c)}
\end{equation}

The vacuum eigenvalues of the integrals of motion do not depend directly on the radius $r$, but rather on the momentum $P(r)$, which is itself a function of $r$:

\begin{equation}
	I_1(r) = P(r)^2 - \frac{1}{24} \; , \qquad I_3(r) = I_1(r)^2 + \frac{1}{6}I_1(r) + \frac{c}{1440} \; .
\end{equation}
As explained neatly in Ref. \cite{AZam_AlZa_96}, in the limit $r\rightarrow 0$ the main contribution to the one-point functions $\langle e^{a\eta}\rangle$, with $a>0$, comes from the following region in the configuration space

\begin{equation}
	\vert b \eta_0 \vert < -\log \frac{\mu^2}{\sin\pi b^2} \; ,
\label{eq:quantizationregion}
\end{equation}
where $\eta_0$ is the zero mode of the field $\eta(z,\bz)$; here the interaction term in sinh-Gordon action can be neglected. This means that in this region we can consider $\eta$ as a free field and that the ground state wave functional $\boldsymbol{\Psi}_0[\eta]$ can be approximated by the superposition of two zero-mode plane waves

\begin{equation}
	\boldsymbol{\Psi}_0[\eta] \underset{r\rightarrow \infty}{\sim} \left( c_1 e^{i P(r) \eta_0} + c_2 e^{-i P(r) \eta_0}\right) \; ,
\end{equation}
where the momentum $P(r)$ is quantised thanks to the presence of the potential walls $b\eta_0 \sim \pm \log \frac{\mu^2}{\sin\pi b^2}$. The quantisation condition reads

\begin{equation}
	S(P)^2 = 1 \; \Rightarrow \; \delta(P) = \pi \; , \ S(P)\doteq e^{-i\delta(P)} \; ,
\end{equation}
where $S(P)$ is the Liouville reflection amplitude

\begin{equation}
	S(P) = -\left(\bmu\frac{\Gamma(1+b^2)}{b^2}\right)^{-4 i \frac{P(r)}{b}}\frac{\Gamma\big(1+2 i P(r) b\big)\Gamma\big(1+2 i P(r) b^{-1}\big)}{\Gamma\big(1-2 i P(r) b\big)\Gamma\big(1-2 i P(r) b^{-1}\big)} \; .
\end{equation}
Using (\ref{eq:massformula}) and remembering that we rescaled the mass $m\rightarrow m R$, we easily obtain the quantisation condition for the momentum

\begin{align}
	&2 P(r) Q \log\left[\frac{r}{8 \pi^{\frac{3}{2}} \big(b^2\big)^{\frac{1}{1+b^2}}}\Gamma\Big(\frac{1}{2(1+b^2)}\Big)\Gamma\Big(1+\frac{b^2}{2(1+b^2)}\Big)\right] = \nonumber
	\\	\label{eq:quantcond}
	\\
	&= -\frac{\pi}{2} + \frac{1}{2 i}\log\left[\frac{\Gamma\big(1+2 i P(r) b\big)\Gamma\big(1+2 i P(r) b^{-1}\big)}{\Gamma\big(1-2 i P(r) b\big)\Gamma\big(1-2 i P(r) b^{-1}\big)}\right] \; .	\nonumber
\end{align}

\begin{figure}[!h]
\centerline{\includegraphics[scale=1]{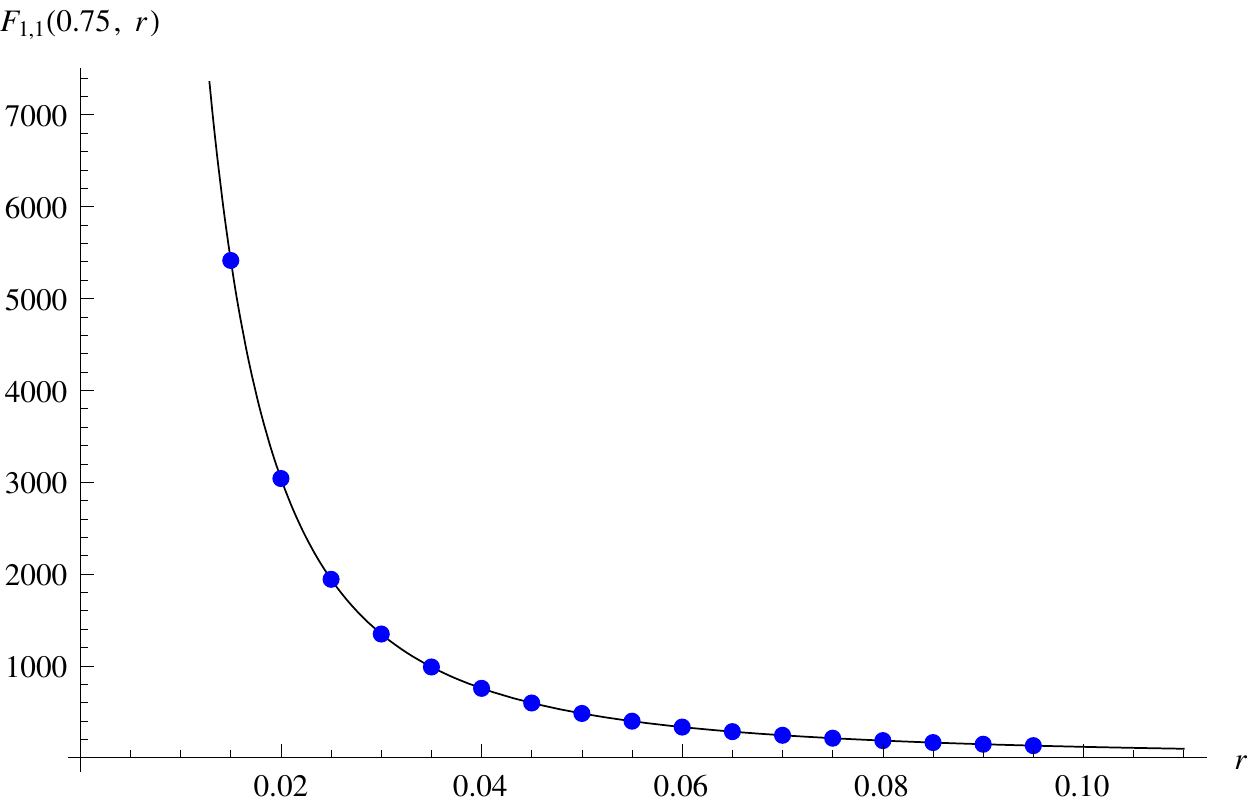}}
\caption{Plot of $F_{1,1}(\alpha,r)$ against its theoretical behaviour for $\alpha =0.75$ and $b=0.4$ \label{fig:f11_a075_b4}}
\end{figure}

\begin{figure}[!h]
\centerline{\includegraphics[scale=1]{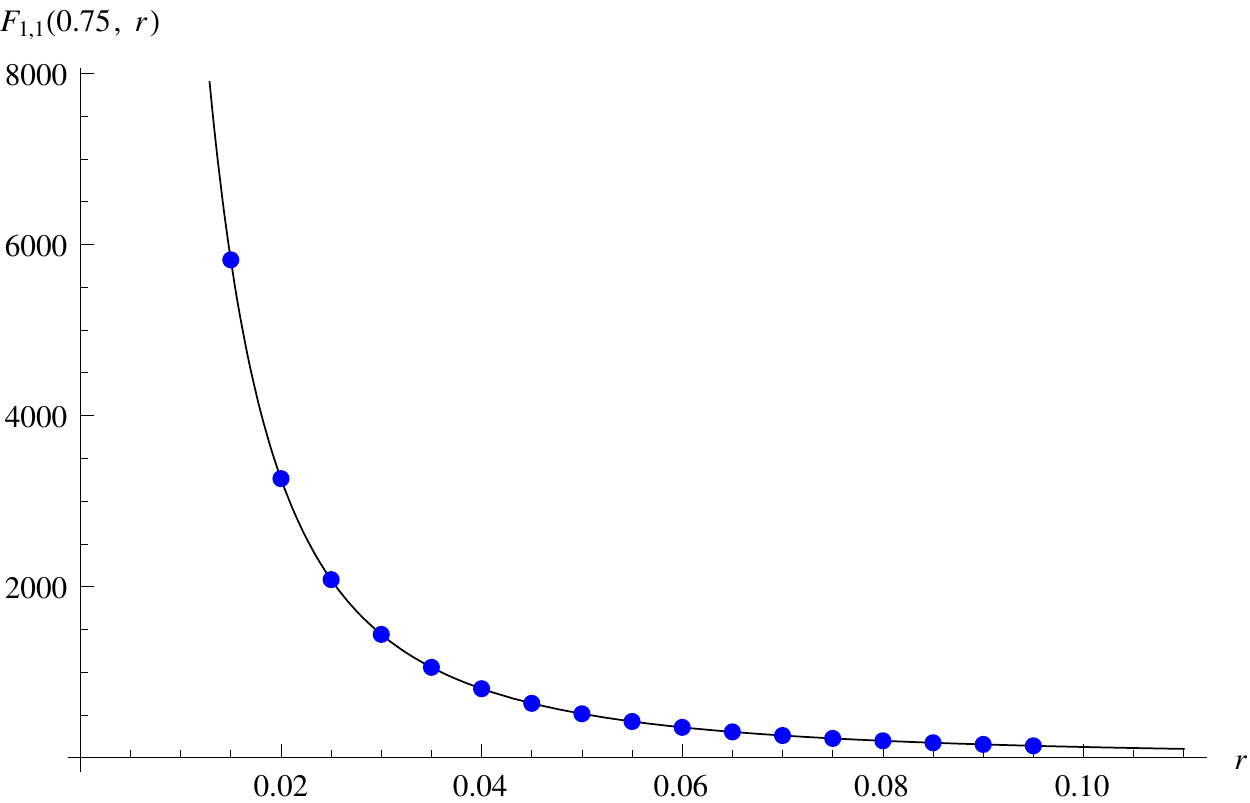}}
\caption{Plot of $F_{1,1}(\alpha,r)$ against its theoretical behaviour for $\alpha =0.75$ and $b=0.8$ \label{fig:f11_a075_b8}}
\end{figure}

\begin{figure}[!h]
\centerline{\includegraphics[scale=1]{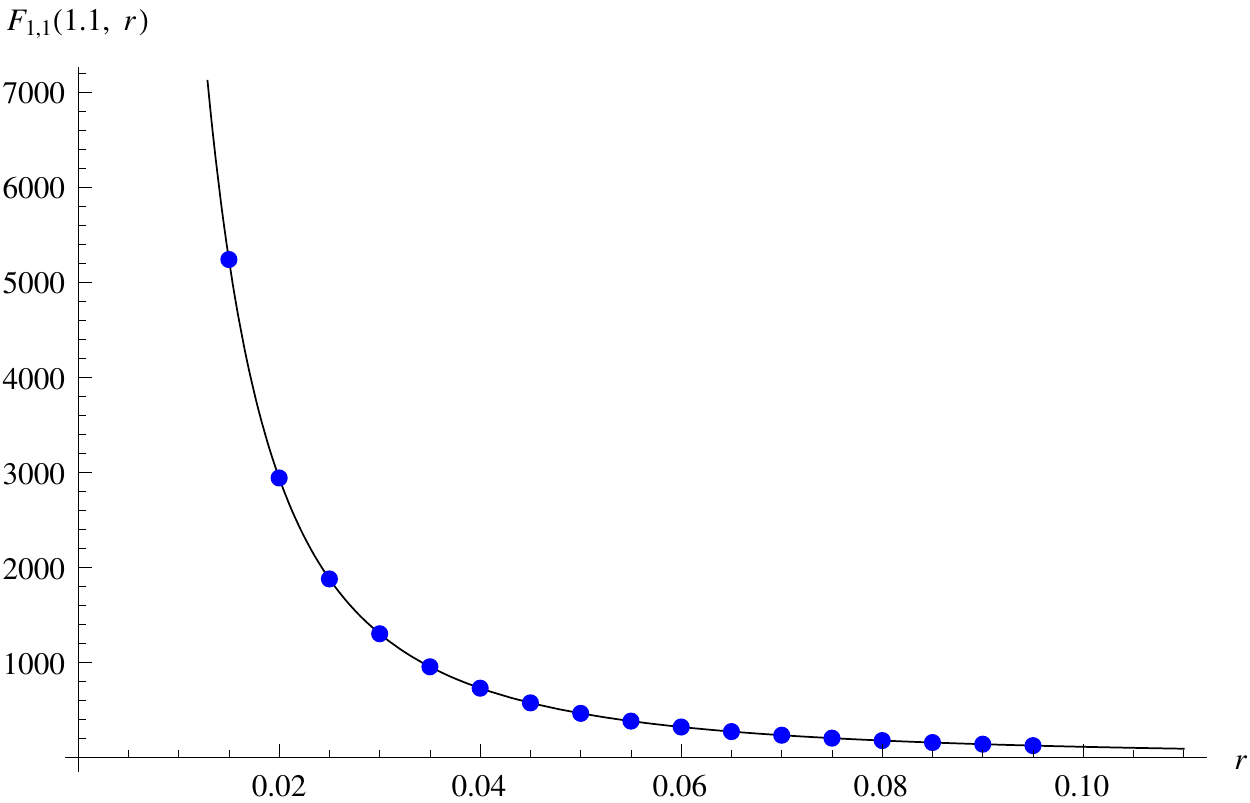}}
\caption{Plot of $F_{1,1}(\alpha,r)$ against its theoretical behaviour for $\alpha =1.1$ and $b=0.4$ \label{fig:f11_a11_b4}}
\end{figure}

\begin{figure}[!h]
\centerline{\includegraphics[scale=1]{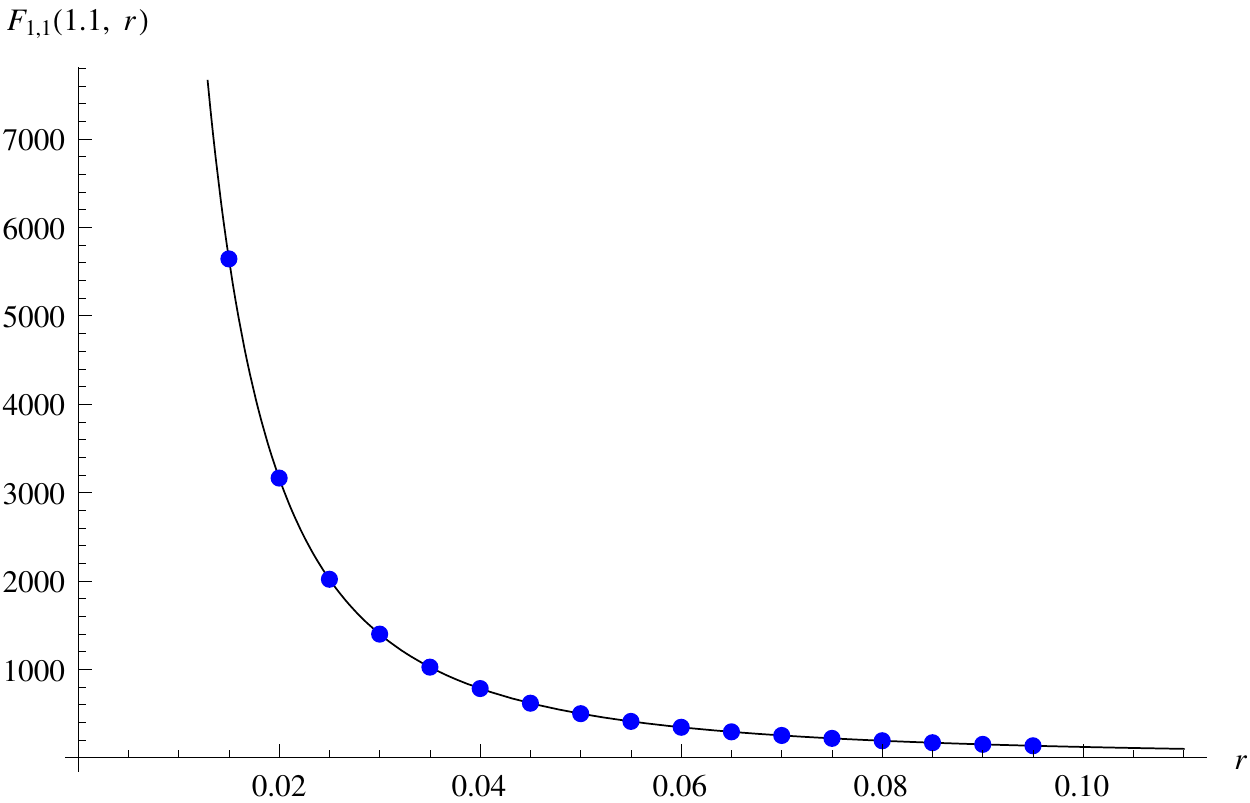}}
\caption{Plot of $F_{1,1}(\alpha,r)$ against its theoretical behaviour for $\alpha =1.1$ and $b=0.8$ \label{fig:f11_a11_b8}}
\end{figure}

\begin{figure}[!h]
\centerline{\includegraphics[scale=1]{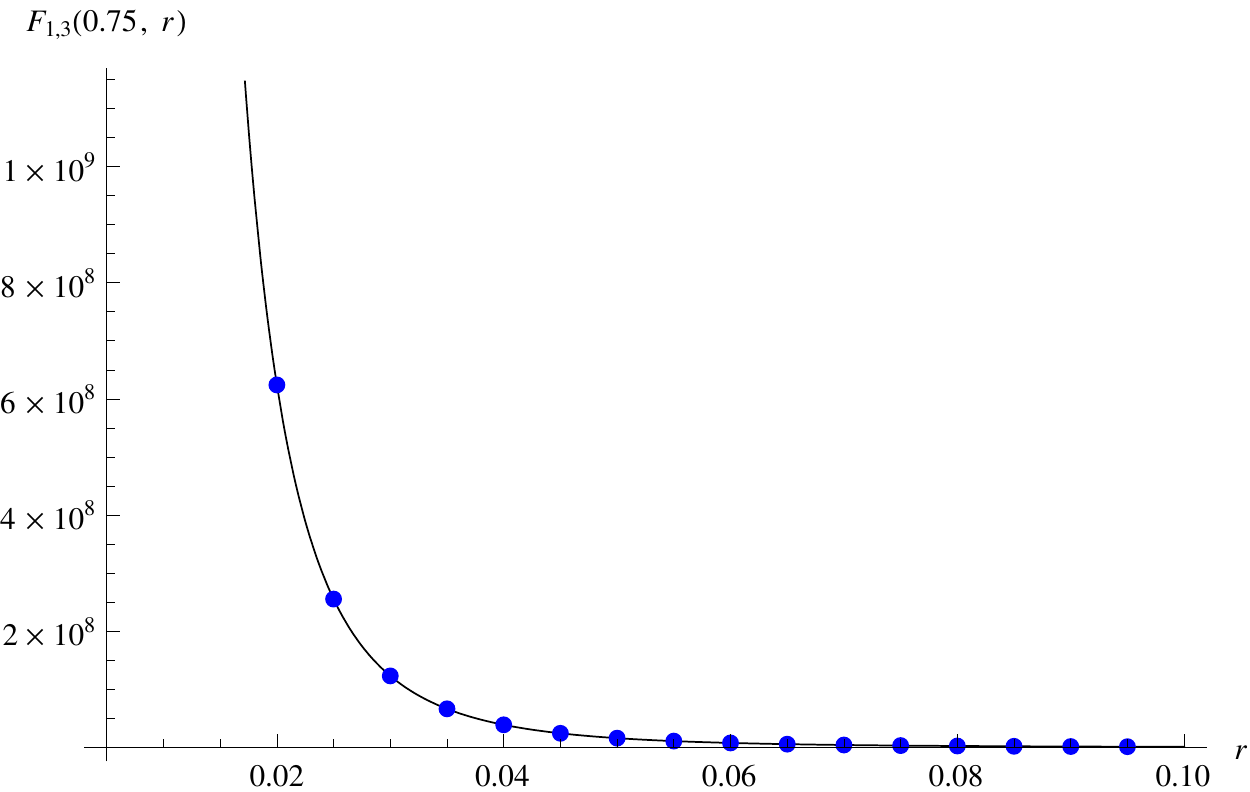}}
\caption{Plot of $F_{1,3}(\alpha,r)$ against its theoretical behaviour for $\alpha =0.75$ and $b=0.4$ \label{fig:f13_a075_b4}}
\end{figure}

\begin{figure}[!h]
\centerline{\includegraphics[scale=1]{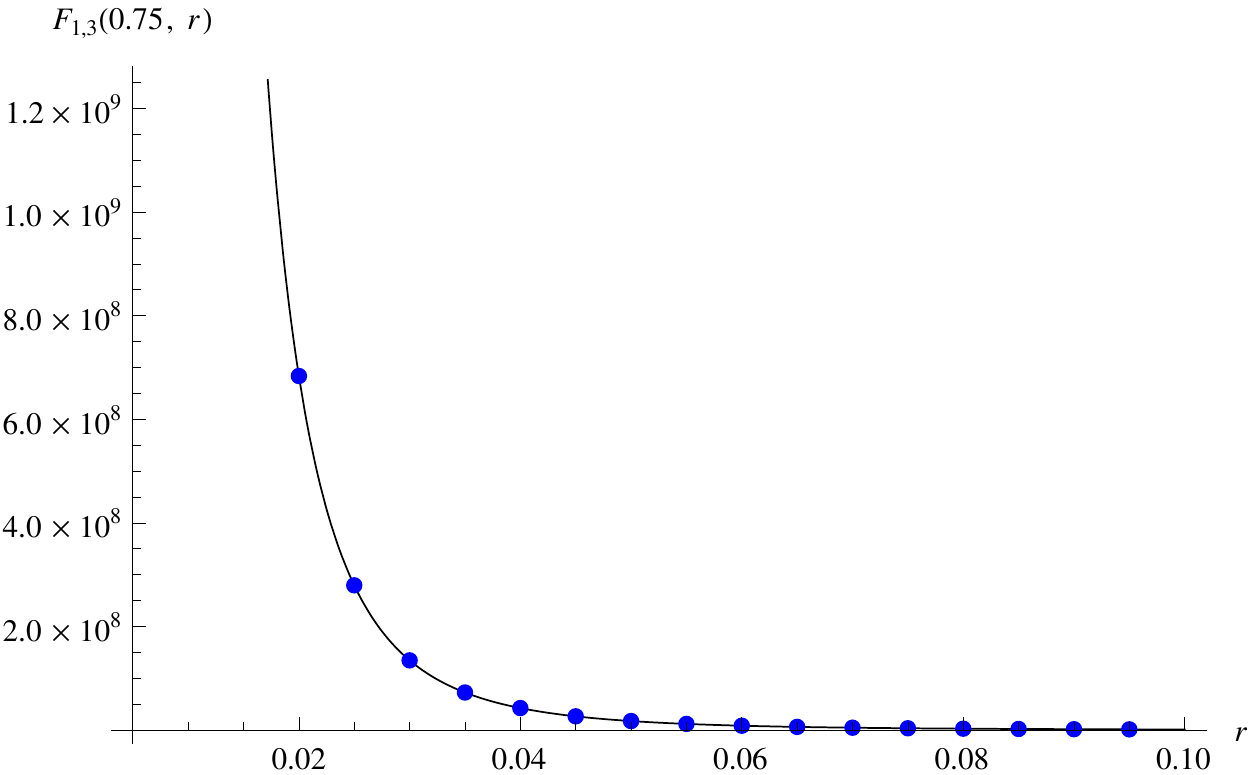}}
\caption{Plot of $F_{1,3}(\alpha,r)$ against its theoretical behaviour for $\alpha =0.75$ and $b=0.8$ \label{fig:f13_a075_b8}}
\end{figure}

\begin{figure}[!h]
\centerline{\includegraphics[scale=1]{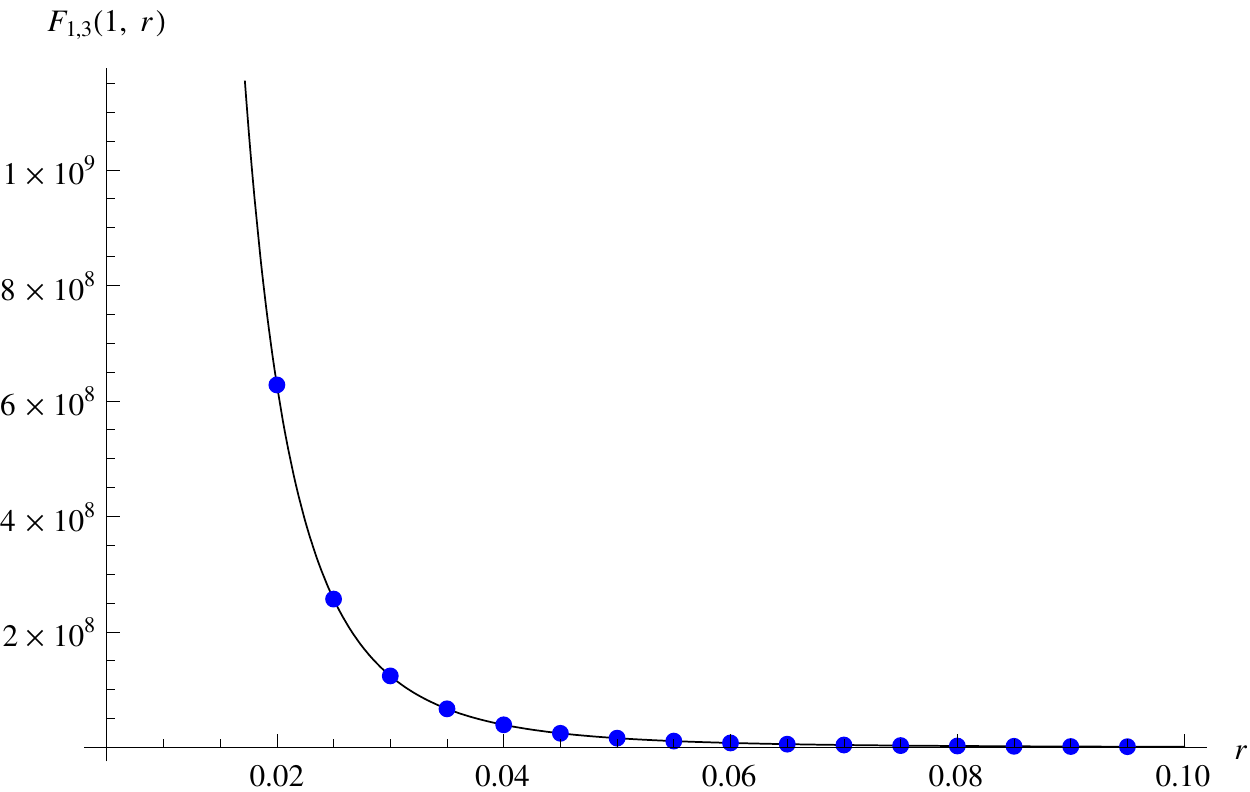}}
\caption{Plot of $F_{1,3}(\alpha,r)$ against its theoretical behaviour for $\alpha =1$ and $b=0.4$ \label{fig:f13_a11_b4}}
\end{figure}

\begin{figure}[!h]
\centerline{\includegraphics[scale=1]{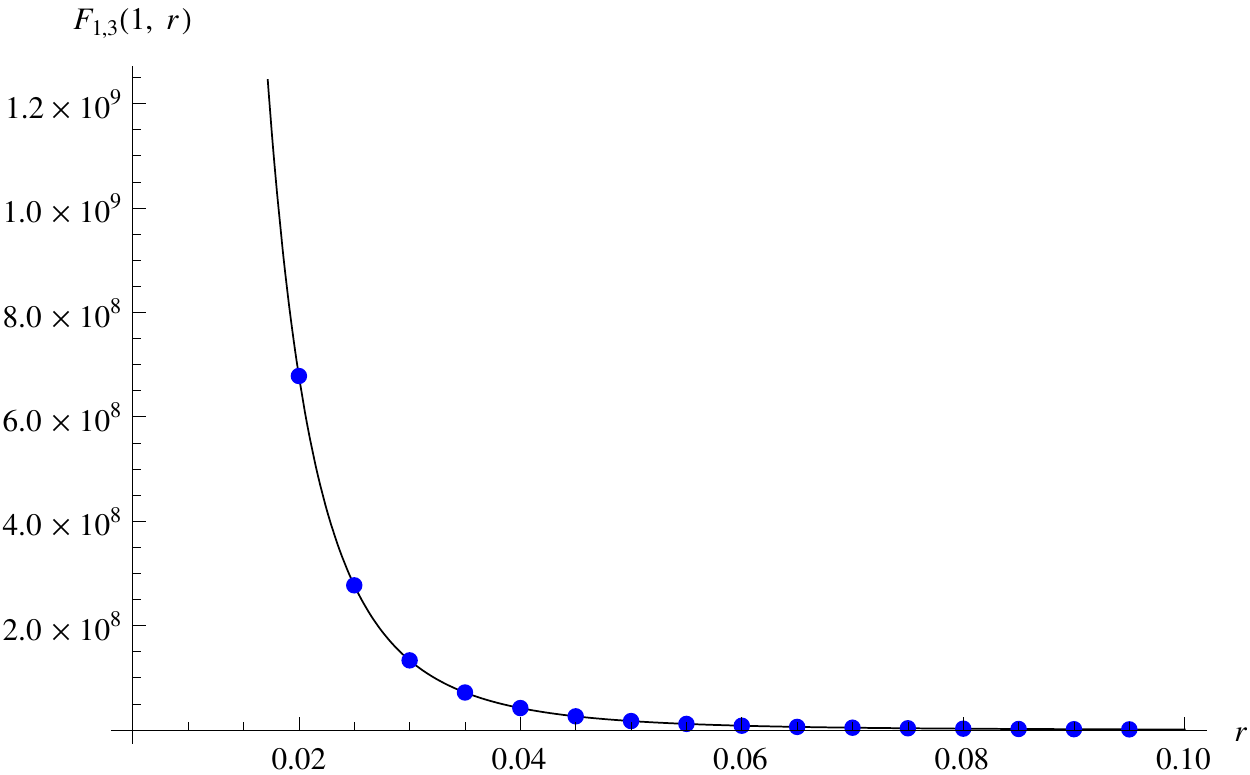}}
\caption{Plot of $F_{1,3}(\alpha,r)$ against its theoretical behaviour for $\alpha =1$ and $b=0.8$ \label{fig:f13_a11_b8}}
\end{figure}

We have considered the two following ratios of expectation values

\begin{equation}
	F_{1,1}(\alpha,r) \doteq \frac{\langle\bbeta^\ast_1\bgamma^\ast_1\Phi_\alpha(0)\rangle_r}{\langle\Phi_\alpha(0)\rangle_r} \; , \qquad	F_{1,3}(\alpha,r) \doteq \frac{\langle\bbeta^\ast_1\bgamma^\ast_3\Phi_\alpha(0)\rangle_r}{\langle\Phi_\alpha(0)\rangle_r} \; ,
\end{equation}
and evaluated numerically the corresponding functions $\Theta_r^{\textrm{shG}}(i,i\vert\alpha)$ and $\Theta_r^{\textrm{shG}}(i,3i\vert\alpha)$ for values of $\alpha$ ranging from $0.75$ up to $1.5$, with $b\in [0.4,1.0]$ and $r\in [0.005,0.95]$. Figures \ref{fig:f11_a075_b4}-\ref{fig:f13_a11_b8} show some of these numerical estimates plotted against the curve (\ref{eq:theodesc}); the agreement of the data with the theoretical prevision is very good for the whole range of $r$ considered.

The tables \ref{tab:f11relerr} and Tab.\ref{tab:f13relerr}, displaying the values of the relative error $\sigma$

\begin{equation}
	\sigma_{2j-1,2k-1} \doteq \left\vert 1-\frac{F_{2j-1,2k-1}(\alpha,r)}{F_{2j-1,2k-1}^{\textrm{CFT}}(\alpha,r)}\right\vert
\end{equation}
with

\begin{equation}
 F_{2j-1,2k-1}^{\textrm{CFT}}(\alpha ,r) = -\left(\frac{2\pi m}{r}\right)^{2j+2k-2} \frac{D_{2j-1}(\alpha)D_{2k-1}(2-\alpha)}{j+k-1}\Omega_{2j-1,2k-1} \; ,
\end{equation}
are a remarkable evidence in support of the conjecture \ref{conj:conj}.

\begin{table}[!h]
\caption{Values of the relative error for $F_{1,1}(\alpha,r)$.}
{\begin{tabular}{l | c | c |c| c | c |}
\cline{2-6}
 & \multicolumn{5}{c|}{$\sigma_{1,1}$}\\
 \cline{2-6}
 & \multicolumn{2}{c|}{$\alpha = 0.75$} & & \multicolumn{2}{c|}{$\alpha = 1.1$}\\
 \hline
 \multicolumn{1}{|c|}{$r$} & $b=0.4$ & $b=0.8$ & & $b=0.4$ & $b=0.8$\\
 \hline
 \multicolumn{1}{|l|}{0.005} & $1.5 \times 10^{-4}$  & $2.0 \times 10^{-5}$ & &  $2.4 \times 10^{-4}$  & $6.0 \times 10^{-5}$ \\
 \multicolumn{1}{|l|}{0.01}  & $5.5 \times 10^{-5}$  & $3.3 \times 10^{-6}$ & &  $1.1 \times 10^{-4}$  & $1.5 \times 10^{-5}$ \\
 \multicolumn{1}{|l|}{0.015} & $2.3 \times 10^{-5}$  & $1.2 \times 10^{-6}$ & &  $6.1 \times 10^{-5}$  & $8.1 \times 10^{-6}$ \\
 \multicolumn{1}{|l|}{0.02}  & $1.3 \times 10^{-5}$  & $1.8 \times 10^{-6}$ & &  $3.7 \times 10^{-5}$  & $4.6 \times 10^{-6}$ \\
 \multicolumn{1}{|l|}{0.025} & $7.5 \times 10^{-6}$  & $3.0 \times 10^{-7}$ & &  $2.2 \times 10^{-5}$  & $2.3 \times 10^{-6}$ \\
 \multicolumn{1}{|l|}{0.03}  & $5.1 \times 10^{-6}$  & $7.3 \times 10^{-7}$ & &  $1.6 \times 10^{-5}$  & $2.8 \times 10^{-6}$ \\
 \multicolumn{1}{|l|}{0.035} & $1.7 \times 10^{-6}$  & $1.1 \times 10^{-6}$ & &  $1.1 \times 10^{-5}$  & $1.1 \times 10^{-6}$ \\
 \multicolumn{1}{|l|}{0.04}  & $1.4 \times 10^{-6}$  & $1.1 \times 10^{-6}$ & &  $7.0 \times 10^{-6}$  & $3.1 \times 10^{-7}$ \\
 \multicolumn{1}{|l|}{0.045} & $1.4 \times 10^{-6}$  & $1.1 \times 10^{-6}$ & &  $6.7 \times 10^{-6}$  & $2.2 \times 10^{-6}$ \\
 \multicolumn{1}{|l|}{0.05}  & $1.3 \times 10^{-6}$  & $1.3 \times 10^{-6}$ & &  $2.4 \times 10^{-6}$  & $2.5 \times 10^{-6}$ \\
 \multicolumn{1}{|l|}{0.055} & $4.2 \times 10^{-6}$  & $3.3 \times 10^{-6}$ & &  $7.1 \times 10^{-6}$  & $8.0 \times 10^{-7}$ \\
 \multicolumn{1}{|l|}{0.06}  & $1.1 \times 10^{-6}$  & $2.5 \times 10^{-6}$ & &  $2.2 \times 10^{-6}$  & $3.2 \times 10^{-7}$ \\
 \multicolumn{1}{|l|}{0.065} & $4.0 \times 10^{-7}$  & $2.4 \times 10^{-7}$ & &  $2.4 \times 10^{-6}$  & $4.4 \times 10^{-7}$ \\
 \multicolumn{1}{|l|}{0.07}  & $2.7 \times 10^{-7}$  & $2.9 \times 10^{-7}$ & &  $2.2 \times 10^{-6}$  & $1.7 \times 10^{-7}$ \\
 \multicolumn{1}{|l|}{0.075} & $3.1 \times 10^{-7}$  & $2.8 \times 10^{-7}$ & &  $1.3 \times 10^{-6}$  & $1.2 \times 10^{-6}$ \\
 \multicolumn{1}{|l|}{0.08}  & $1.3 \times 10^{-7}$  & $1.0 \times 10^{-7}$ & &  $1.0 \times 10^{-6}$  & $4.3 \times 10^{-8}$ \\
 \multicolumn{1}{|l|}{0.085} & $5.4 \times 10^{-7}$  & $1.3 \times 10^{-7}$ & &  $3.1 \times 10^{-8}$  & $5.4 \times 10^{-7}$ \\
 \multicolumn{1}{|l|}{0.09}  & $2.8 \times 10^{-8}$  & $2.2 \times 10^{-7}$ & &  $1.4 \times 10^{-6}$  & $2.8 \times 10^{-6}$ \\
 \multicolumn{1}{|l|}{0.095} & $1.2 \times 10^{-6}$  & $2.3 \times 10^{-7}$ & &  $2.6 \times 10^{-6}$  & $6.8 \times 10^{-11}$ \\
 \hline
\end{tabular} \label{tab:f11relerr}}
\end{table}

\begin{table}[!h]
\caption{Values of the relative error for $F_{1,3}(\alpha,r)$.}
{\begin{tabular}{l | c | c |c| c | c |}
\cline{2-6}
 & \multicolumn{5}{c|}{$\sigma_{1,3}$}\\
 \cline{2-6}
 & \multicolumn{2}{c|}{$\alpha = 0.75$} & & \multicolumn{2}{c|}{$\alpha = 1$}\\
 \hline
 \multicolumn{1}{|c|}{$r$} & $b=0.4$ & $b=0.8$ & & $b=0.4$ & $b=0.8$\\
 \hline
 \multicolumn{1}{|l|}{0.005} & $2.4 \times 10^{-4}$  & $6.2 \times 10^{-5}$ & &  $1.4 \times 10^{-4}$  & $1.8 \times 10^{-5}$ \\
 \multicolumn{1}{|l|}{0.01}  & $1.0 \times 10^{-4}$  & $1.6 \times 10^{-5}$ & &  $5.2 \times 10^{-5}$  & $5.0 \times 10^{-6}$ \\
 \multicolumn{1}{|l|}{0.015} & $6.2 \times 10^{-5}$  & $9.3 \times 10^{-6}$ & &  $2.5 \times 10^{-5}$  & $3.1 \times 10^{-6}$ \\
 \multicolumn{1}{|l|}{0.02}  & $3.7 \times 10^{-5}$  & $4.1 \times 10^{-6}$ & &  $1.2 \times 10^{-5}$  & $1.1 \times 10^{-6}$ \\
 \multicolumn{1}{|l|}{0.025} & $2.1 \times 10^{-5}$  & $8.9 \times 10^{-7}$ & &  $7.0 \times 10^{-6}$  & $5.0 \times 10^{-7}$ \\
 \multicolumn{1}{|l|}{0.03}  & $1.8 \times 10^{-5}$  & $8.1 \times 10^{-7}$ & &  $4.2 \times 10^{-6}$  & $8.8 \times 10^{-7}$ \\
 \multicolumn{1}{|l|}{0.035} & $1.0 \times 10^{-5}$  & $8.5 \times 10^{-7}$ & &  $1.7 \times 10^{-6}$  & $9.9 \times 10^{-7}$ \\
 \multicolumn{1}{|l|}{0.04}  & $7.4 \times 10^{-6}$  & $1.4 \times 10^{-6}$ & &  $1.0 \times 10^{-6}$  & $6.7 \times 10^{-7}$ \\
 \multicolumn{1}{|l|}{0.045} & $5.9 \times 10^{-6}$  & $1.3 \times 10^{-6}$ & &  $4.1 \times 10^{-7}$  & $2.1 \times 10^{-6}$ \\
 \multicolumn{1}{|l|}{0.05}  & $3.3 \times 10^{-6}$  & $2.2 \times 10^{-6}$ & &  $3.4 \times 10^{-7}$  & $1.0 \times 10^{-6}$ \\
 \multicolumn{1}{|l|}{0.055} & $2.4 \times 10^{-6}$  & $1.2 \times 10^{-6}$ & &  $1.3 \times 10^{-6}$  & $2.2 \times 10^{-6}$ \\
 \multicolumn{1}{|l|}{0.06}  & $2.2 \times 10^{-6}$  & $4.8 \times 10^{-7}$ & &  $7.1 \times 10^{-7}$  & $8.0 \times 10^{-7}$ \\
 \multicolumn{1}{|l|}{0.065} & $1.6 \times 10^{-6}$  & $7.0 \times 10^{-7}$ & &  $3.9 \times 10^{-7}$  & $4.2 \times 10^{-8}$ \\
 \multicolumn{1}{|l|}{0.07}  & $1.3 \times 10^{-8}$  & $4.9 \times 10^{-7}$ & &  $3.0 \times 10^{-7}$  & $4.7 \times 10^{-7}$ \\
 \multicolumn{1}{|l|}{0.075} & $3.8 \times 10^{-7}$  & $1.1 \times 10^{-6}$ & &  $7.6 \times 10^{-9}$  & $5.8 \times 10^{-7}$ \\
 \multicolumn{1}{|l|}{0.08}  & $8.0 \times 10^{-7}$  & $1.2 \times 10^{-6}$ & &  $9.2 \times 10^{-8}$  & $3.6 \times 10^{-7}$ \\
 \multicolumn{1}{|l|}{0.085} & $2.8 \times 10^{-6}$  & $4.2 \times 10^{-7}$ & &  $4.5 \times 10^{-7}$  & $6.1 \times 10^{-7}$ \\
 \multicolumn{1}{|l|}{0.09}  & $1.6 \times 10^{-6}$  & $2.8 \times 10^{-6}$ & &  $7.0 \times 10^{-7}$  & $4.0 \times 10^{-7}$ \\
 \multicolumn{1}{|l|}{0.095} & $3.0 \times 10^{-6}$  & $1.0 \times 10^{-6}$ & &  $7.1 \times 10^{-7}$  & $6.1 \times 10^{-7}$ \\
 \hline
\end{tabular} \label{tab:f13relerr}}
\end{table}

\subsection{Primary fields}
\label{subsec:primary}

Let us now consider the following ratio of primary fields' expectation values

\begin{equation}
	\mathcal F(\alpha,r) \doteq \frac{\langle\Phi_{\alpha -2\frac{b^2}{b^2+1}}\rangle^{\textrm{shG}}_r}{\langle\Phi_\alpha\rangle^{\textrm{shG}}_r}\; .
\end{equation}
Using the shift formula (\ref{eq:shiftformula}) and the determinant one (\ref{eq:onepointfunctionsinh}) we can write

\begin{equation}
	\mathcal F(\alpha,r) = \frac{C_1(\alpha)}{t_1(\alpha)} \frac{\langle\bbeta^\ast_1\bar\bgamma^\ast_1\Phi_{\alpha}\rangle^{\textrm{shG}}_r}{\langle\Phi_\alpha\rangle^{\textrm{shG}}_r} = - \frac{C_1(\alpha)}{\pi t_1(\alpha)}\Big[\Theta(i,-i\vert\alpha) - \pi t_1(\alpha)\Big] \; .
\label{eq:finthetaterms}
\end{equation}

On the other hand, from Ref. \cite{Luky_01} we know that we can approximate the behaviour of the expectation value of a primary field $\Phi_\alpha$ in the region (\ref{eq:quantizationregion}) with that of a three-point function of Liouville CFT:

\begin{equation}
	\langle\Phi_\alpha\rangle^{\textrm{shG}}_r \underset{r\rightarrow 0}{\sim} \mathcal N(r,b) \langle 0\vert e^{a(-P)\eta(-\infty)} \Phi_\alpha e^{a(P)\eta(\infty)}\vert 0\rangle_r^{\textrm{Liou}}
\end{equation}
where the function $\mathcal N(r,b)$ is a normalization constant and

\begin{equation}
	a(P) = \frac{Q}{2} + i P(r) \; \Rightarrow \qquad \Delta_{a(P)} = \frac{Q^2}{4} - P(r)^2
\end{equation}
with $P(r)$ satisfying the quantization condition (\ref{eq:quantcond}).

The form of Liouville three-point function was found in Refs. \cite{Dorn_Otto_94} and \cite{AZam_AlZa_96} and reads

\begin{equation}
	\langle 0\vert e^{a(-P)\eta(-\infty)} \Phi_\alpha e^{a(P)\eta(\infty)}\vert 0\rangle_r^{\textrm{Liou}} = \left(\bmu \frac{\Gamma(1+b^2)}{b^{1+b^2}}\right)^{-Q\frac{\alpha}{b}} \Upsilon_0 \frac{\Upsilon(2a)\Upsilon(Q-2 i P)\Upsilon(Q+2 i P)}{\Upsilon(a)^2\Upsilon(a-2 i P)\Upsilon(a+2 i P)} \; ,
\label{eq:liouvillethreepoint}
\end{equation}
where the function $\Upsilon(x)$ is defined by the equations

\begin{equation}
	\frac{\Upsilon(x+b)}{\Upsilon(x)} = \gamma(b\,x) b^{1-2bx} \; , \quad \frac{\Upsilon(x+b^{-1})}{\Upsilon(x)} = \gamma\left(\frac{x}{b}\right) b^{-1+2\frac{x}{b}} \; , \quad \Upsilon_0 \doteq \frac{d \Upsilon}{dx}\Big\vert_{x=0} \; .	\nonumber
\end{equation}
The general form of the normalization $\mathcal N(r,b)$ is not known, but this is irrelevant to our needs, since we are considering the ratio of two one-point functions.

With some simple calculations one finds

\begin{align}
	\mathcal F(\alpha,r) &\underset{r\rightarrow 0}{\sim} \mathcal F^{\textrm{CFT}}(\alpha,r) = \left[\frac{r}{8\pi^{\frac{3}{2}}}\Gamma\left(\frac{1}{2(1+b^2)}\right)\Gamma\left(1+\frac{b^2}{2(1+b^2)}\right)\right]^2 \times	\nonumber
	\\
	&\times \frac{\gamma\big(b(a-b)\big)^2}{\gamma\big(b(2a-b)\big)\gamma\big(2b(a-b)\big)} \gamma\big(b(a-b+2 i P)\big)\gamma\big(b(a-b-2 i P)\big) \; .	\label{eq:fcftbehaviour}
\end{align}

\begin{figure}[!h]
\centerline{\includegraphics[scale=1]{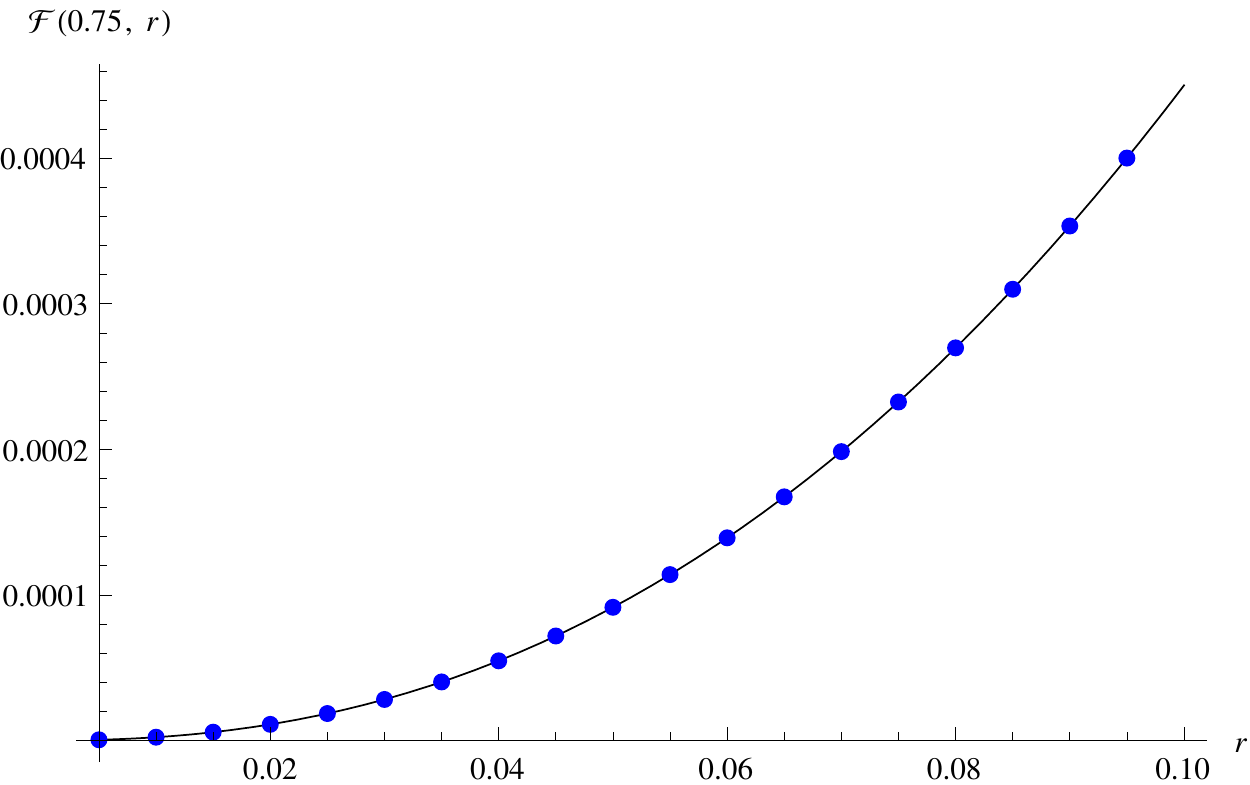}}
\caption{Plot of $\mathcal F(\alpha,r)$ against its theoretical behaviour for $\alpha =0.75$ and $b=0.4$ \label{fig:eff_a075_b4}}
\end{figure}

\begin{figure}[!h]
\centerline{\includegraphics[scale=1]{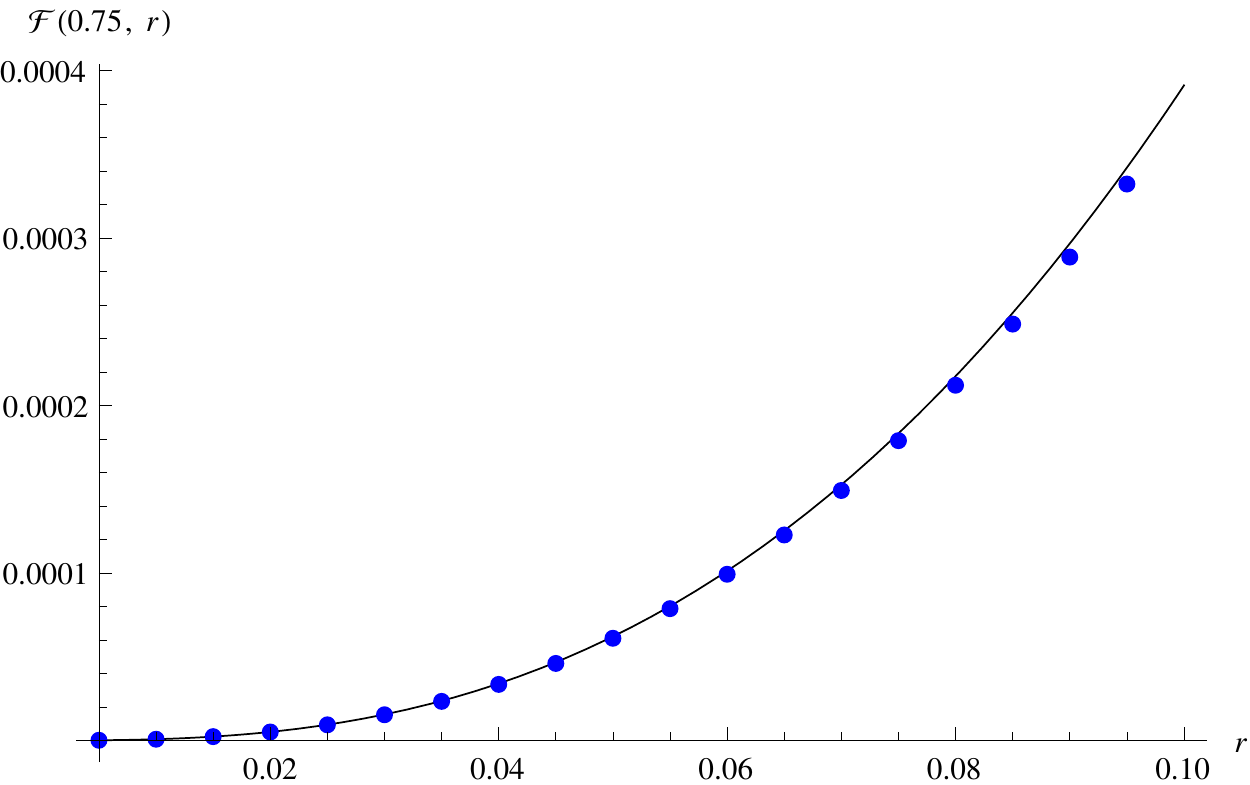}}
\caption{Plot of $\mathcal F(\alpha,r)$ against its theoretical behaviour for $\alpha =0.75$ and $b=0.7$ \label{fig:eff_a075_b7}}
\end{figure}

\begin{figure}[!h]
\centerline{\includegraphics[scale=1]{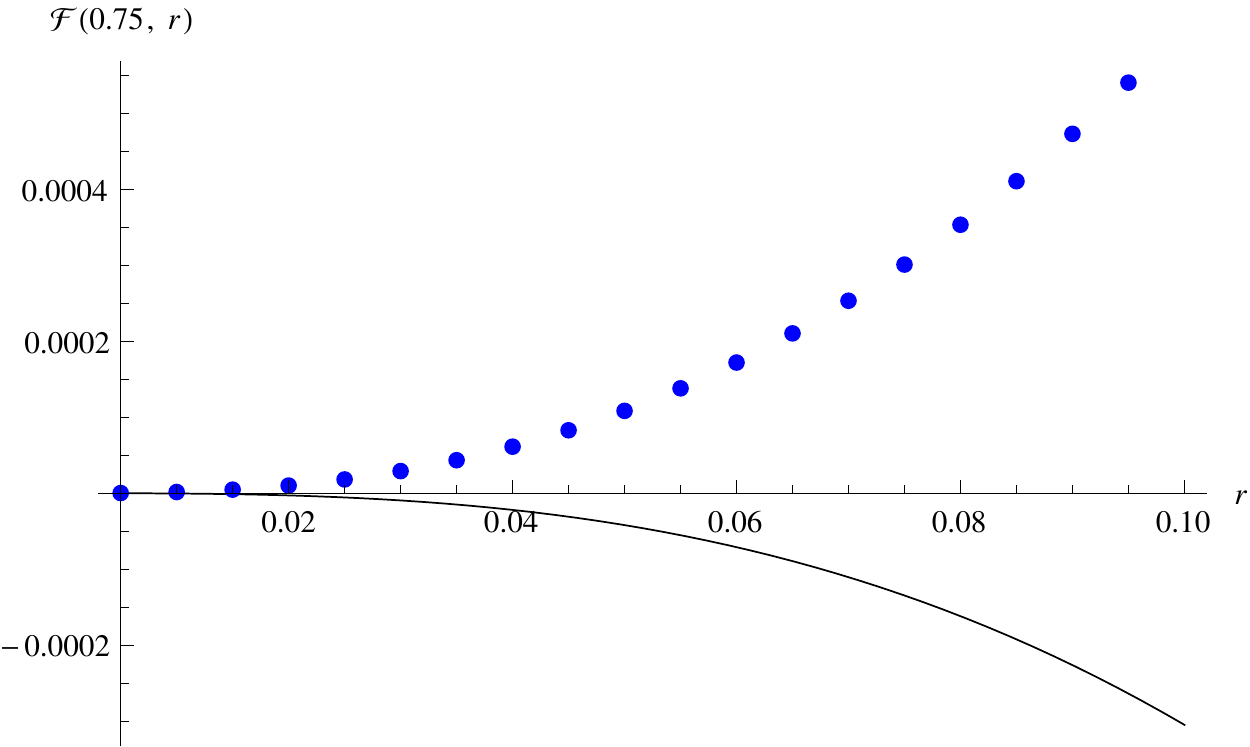}}
\caption{Plot of $\mathcal F(\alpha,r)$ against its theoretical behaviour for $\alpha =0.75$ and $b=0.8$ \label{fig:eff_a075_b8}}
\end{figure}

\begin{figure}[!h]
\centerline{\includegraphics[scale=1]{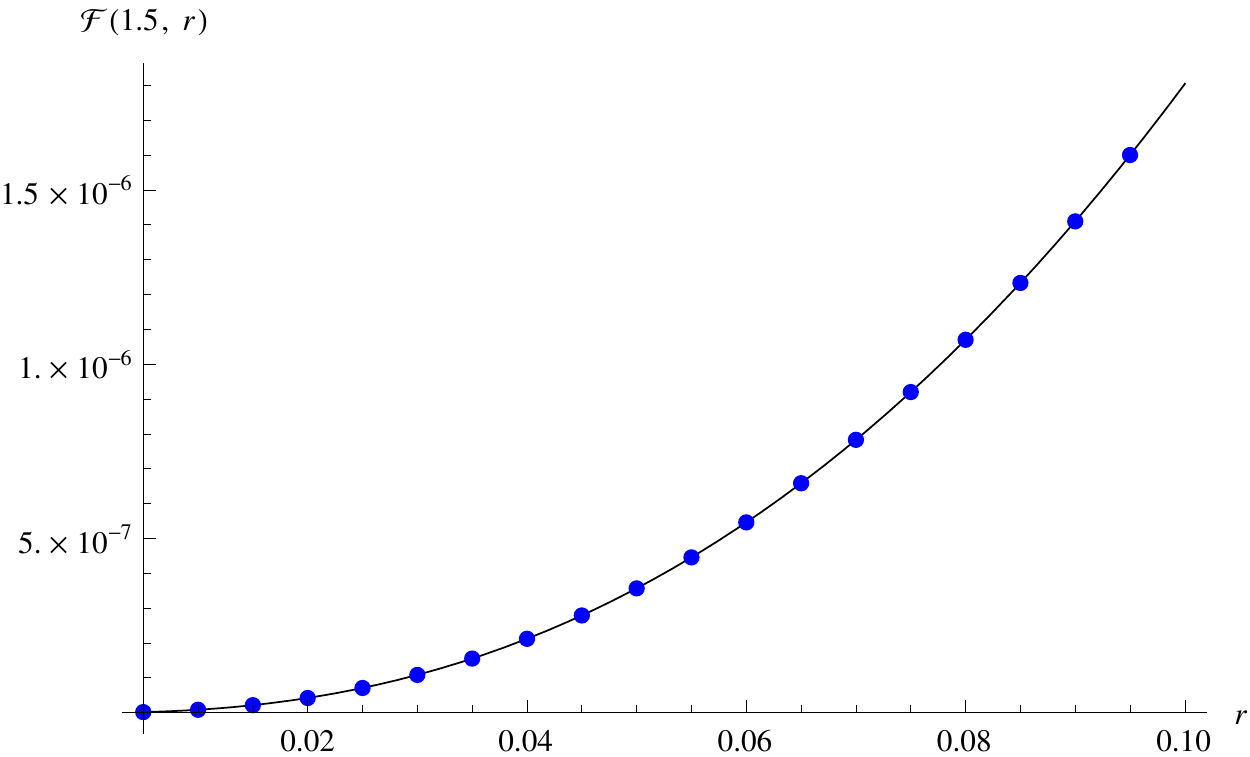}}
\caption{Plot of $\mathcal F(\alpha,r)$ against its theoretical behaviour for $\alpha =1.5$ and $b=0.4$ \label{fig:eff_a15_b4}}
\end{figure}

\begin{figure}[!h]
\centerline{\includegraphics[scale=1]{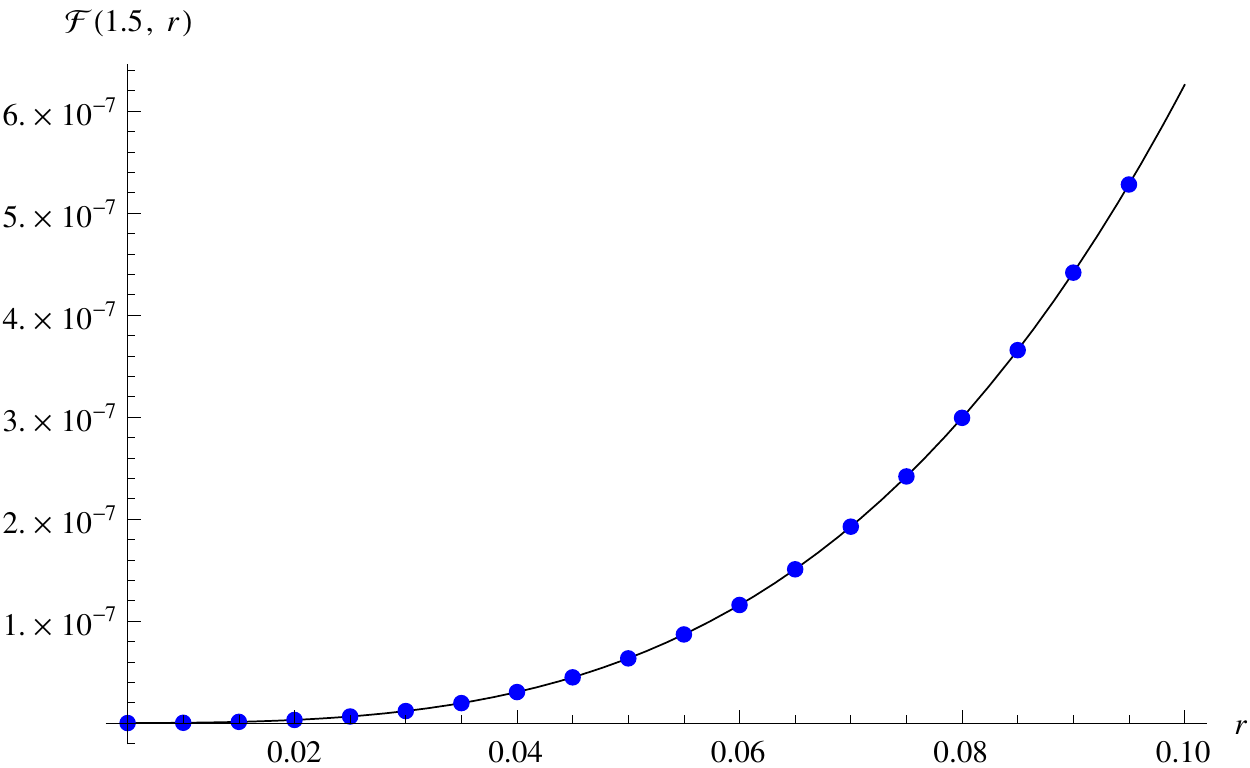}}
\caption{Plot of $\mathcal F(\alpha,r)$ against its theoretical behaviour for $\alpha =1.5$ and $b=0.8$ \label{fig:eff_a15_b8}}
\end{figure}

We have evaluated numerically the function $\Theta(i,-i\vert\alpha)$ and used it to extract the value of $\mathcal F(\alpha,r)$ by means of the formula (\ref{eq:finthetaterms}). We then compared the data we obtained with the theoretical CFT behaviour (\ref{eq:fcftbehaviour}). Figures \ref{fig:eff_a075_b4}-\ref{fig:eff_a15_b8} show the results for various values of $\alpha$ and $b$.

In table \ref{tab:effrelerr} are collected the values of the relative error $\varsigma$

\begin{equation}
	\varsigma \doteq \left\vert 1 - \frac{\mathcal F(\alpha,r)}{\mathcal F^{\textrm{CFT}}(\alpha,r)}\right\vert \; .
\end{equation}

\begin{table}[!h]
\caption{Values of the relative error for $\mathcal F(\alpha,r)$.}
{\begin{tabular}{l | c | c | c |c| c | c |}
\cline{2-7}
 & \multicolumn{6}{c|}{$\varsigma$}\\
 \cline{2-7}
 & \multicolumn{3}{c|}{$\alpha = 0.75$} & & \multicolumn{2}{c|}{$\alpha = 1.5$}\\
 \hline
 \multicolumn{1}{|c|}{$r$} & $b=0.4$ & $b=0.7$ & $b=0.8$ & & $b=0.4$ & $b=0.8$\\
 \hline
 \multicolumn{1}{|l|}{0.005} & $6.2 \times 10^{-3}$ & $1.1 \times 10^{-3}$  & $1.2$ & &  $1.1 \times 10^{-2}$  & $8.2 \times 10^{-4}$ \\
 \multicolumn{1}{|l|}{0.01}  & $2.7 \times 10^{-3}$ & $1.7 \times 10^{-3}$  & $1.2$ & &  $3.6 \times 10^{-3}$  & $2.5 \times 10^{-4}$ \\
 \multicolumn{1}{|l|}{0.015} & $1.4 \times 10^{-3}$ & $5.9 \times 10^{-3}$  & $1.3$ & &  $1.7 \times 10^{-3}$  & $1.1 \times 10^{-4}$ \\
 \multicolumn{1}{|l|}{0.02}  & $7.8 \times 10^{-4}$ & $7.3 \times 10^{-3}$  & $1.3$ & &  $9.6 \times 10^{-4}$  & $5.4 \times 10^{-5}$ \\
 \multicolumn{1}{|l|}{0.025} & $5.7 \times 10^{-4}$ & $9.4 \times 10^{-3}$  & $1.3$ & &  $5.8 \times 10^{-4}$  & $2.7 \times 10^{-5}$ \\
 \multicolumn{1}{|l|}{0.03}  & $3.1 \times 10^{-4}$ & $1.1 \times 10^{-2}$  & $1.3$ & &  $3.7 \times 10^{-4}$  & $1.6 \times 10^{-5}$ \\
 \multicolumn{1}{|l|}{0.035} & $2.2 \times 10^{-4}$ & $1.2 \times 10^{-2}$  & $1.3$ & &  $2.4 \times 10^{-4}$  & $1.0 \times 10^{-5}$ \\
 \multicolumn{1}{|l|}{0.04}  & $1.7 \times 10^{-4}$ & $1.4 \times 10^{-2}$  & $1.4$ & &  $1.7 \times 10^{-4}$  & $5.3 \times 10^{-6}$ \\
 \multicolumn{1}{|l|}{0.045} & $1.6 \times 10^{-4}$ & $1.6 \times 10^{-2}$  & $1.4$ & &  $1.2 \times 10^{-4}$  & $1.2 \times 10^{-6}$ \\
 \multicolumn{1}{|l|}{0.05}  & $3.6 \times 10^{-5}$ & $1.7 \times 10^{-2}$  & $1.4$ & &  $8.5 \times 10^{-5}$  & $5.9 \times 10^{-6}$ \\
 \multicolumn{1}{|l|}{0.055} & $9.5 \times 10^{-5}$ & $1.9 \times 10^{-2}$  & $1.4$ & &  $5.9 \times 10^{-5}$  & $1.1 \times 10^{-6}$ \\
 \multicolumn{1}{|l|}{0.06}  & $3.6 \times 10^{-5}$ & $2.0 \times 10^{-2}$  & $1.4$ & &  $4.4 \times 10^{-5}$  & $6.6 \times 10^{-7}$ \\
 \multicolumn{1}{|l|}{0.065} & $7.2 \times 10^{-5}$ & $2.1 \times 10^{-2}$  & $1.4$ & &  $3.5 \times 10^{-5}$  & $5.6 \times 10^{-7}$ \\
 \multicolumn{1}{|l|}{0.07}  & $5.5 \times 10^{-5}$ & $2.3 \times 10^{-2}$  & $1.4$ & &  $2.6 \times 10^{-5}$  & $9.8 \times 10^{-7}$ \\
 \multicolumn{1}{|l|}{0.075} & $2.8 \times 10^{-5}$ & $2.4 \times 10^{-2}$  & $1.4$ & &  $1.8 \times 10^{-5}$  & $3.3 \times 10^{-7}$ \\
 \multicolumn{1}{|l|}{0.08}  & $3.1 \times 10^{-5}$ & $2.5 \times 10^{-2}$  & $1.5$ & &  $1.3 \times 10^{-5}$  & $9.2 \times 10^{-8}$ \\
 \multicolumn{1}{|l|}{0.085} & $3.1 \times 10^{-5}$ & $2.7 \times 10^{-2}$  & $1.5$ & &  $9.3 \times 10^{-6}$  & $8.2 \times 10^{-8}$ \\
 \multicolumn{1}{|l|}{0.09}  & $7.5 \times 10^{-6}$ & $2.8 \times 10^{-2}$  & $1.5$ & &  $9.1 \times 10^{-6}$  & $2.0 \times 10^{-7}$ \\
 \multicolumn{1}{|l|}{0.095} & $3.7 \times 10^{-6}$ & $2.9 \times 10^{-2}$  & $1.5$ & &  $4.8 \times 10^{-6}$  & $3.9 \times 10^{-7}$ \\
 \hline
\end{tabular} \label{tab:effrelerr}}
\end{table}

The agreement between the data and the CFT behaviour is incredibly good until $b\gtrsim 0.7$, when $\alpha=0.75$, as is clearly visible from figures \ref{fig:eff_a075_b7} and \ref{fig:eff_a075_b8}. The reason for this discrepancy is that, as we explained in the introduction, the supposition that sinh-Gordon approaches na\"\i vely the Liouville CFT in its UV limit is no longer valid when $b\geq\sqrt{\frac{\alpha}{2-\alpha}}$. When $\alpha=0.75$, the \emph{critical} value is $b^{\textrm{crit}} = \sqrt{3/5} \sim 0.774$, which explains why figure \ref{fig:eff_a075_b7} still shows a good agreement for very small values of $r$, while in figure \ref{fig:eff_a075_b8} we see that the data and the CFT curve behave in radically different ways.

\section{Conclusion}

We investigated numerically the behaviour of the conjecture \ref{conj:conj} in the UV limit $R\rightarrow 0$ of the sinh-Gordon model defined on an infinite cylinder of radius $2\pi R$. We found an extremely good agreement with the theoretical predictions in Ref. \cite{Luky_01}, up to the $4^{\textrm{th}}$ decimal place, in the cases of both primary and descendant fields. In figures \ref{fig:f11_a075_b4}-\ref{fig:eff_a15_b8} and tables \ref{tab:f11relerr}-\ref{tab:effrelerr} part of these results are collected. We consider these, along with the analytical results of Ref. \cite{Negr_Smir_13-2}, as a very strong confirmation of the correctness of the fermionic basis description for the sinh-Gordon model.

We have also verified that the limiting behaviour of the primary fields' expectation values is very well described by that of a particular three point function in Liouville CFT only if the parameters are such that the scaling dimensions of the involved fields are all positive, meaning that $0<b<a<Q$. It would be interesting to study the behaviour of sinh-Gordon model's UV limit outside this region.\\

\section*{Acknowledgments}
I am grateful to F. Smirnov whose valuable advices helped considerably to direct my analysis and organize this work.\\
Research of SN is supported by the People Programme (Marie Curie Actions) of the European Union's Seventh Framework Programme FP7/2007-2013/ under REA Grant Agreement No 317089 (GATIS).\\
This project was partially supported by INFN grant IS FTECP and the
UniTo-SanPaolo research grant Nr TO-Call3-2012-0088.


\begin{thebibliography}{00}
\bibitem{Boos_Jimb_Miwa_Smir_Take_07}
       H. Boos, M. Jimbo, T. Miwa, F. Smirnov and Y. Takeyama,
                \emph{Hidden Grassmann structure in the XXZ model},
                Commun. Math. Phys. \textbf{272} (2007) 263-281.

\bibitem{Boos_Jimb_Miwa_Smir_Take_09}
       H. Boos, M. Jimbo, T. Miwa, F. Smirnov and Y. Takeyama,
                \emph{Hidden Grassmann structure in the XXZ model II: creation operators},
                Commun. Math. Phys. \textbf{286} (2009) 875-932.

\bibitem{Jimb_Miwa_Smir_09}
       M. Jimbo, T. Miwa and F. Smirnov,
                \emph{Hidden Grassmann structure in the XXZ model III: introducing Matsubara direction},
                J. Phys. \textbf{A42} (2009) 304018.

\bibitem{Boos_Jimb_Miwa_Smir_10}
       H. Boos, M. Jimbo, T. Miwa and F. Smirnov,
                \emph{Hidden Grassmann structure in the XXZ model IV: CFT limit},
                Commun. Math. Phys. \textbf{299} (2010) 825-866.
                
\bibitem{Jimb_Miwa_Smir_10}
       M. Jimbo, T. Miwa and F. Smirnov,
                \emph{On one point functions of descendents in Sine-Gordon model},
                arXiv: 0912.0934
                
\bibitem{Jimb_Miwa_Smir_11-1}
       M. Jimbo, T. Miwa and F. Smirnov,
                \emph{Hidden Grassmann structure in the XXZ model V: sine-Gordon model},
                Lett. Math. Phys. \textbf{96} (2011) 325-365.

\bibitem{Jimb_Miwa_Smir_11-2}
       M. Jimbo, T. Miwa and F. Smirnov,
                \emph{Fermionic structure in the sine-Gordon model: form factors and null-vectors},
                Nucl. Phys. \textbf{B852} (2011) 390-440.

\bibitem{Negr_Smir_13-2}
	S. Negro and F. Smirnov,
		\emph{On one-point functions for sinh-Gordon model at finite temperature},
		Nucl. Phys. \textbf{B875} (2013) 166-185, arXiv: 1306.1476.

\bibitem{Byts_Tesc_06}
	A.G. Bytsko and J. Teschner,
		\emph{Quantization of models with non-compact quantum group symmetry. Modular XXZ magnet and lattice sinh-Gordon model},
		J. Phys. \textbf{A39} (2006) 12927-12981.

\bibitem{Tesc_07}
	J. Teschner,
		\emph{On the spectrum of the Sinh-Gordon model in finite volume},
		Nucl. Phys. \textbf{B799} (2008) 403-429, [arXiv:hep-th/0702214].

\bibitem{Fate_Frad_Luky_AZam_AlZa_99}
	V. Fateev, D. Fradkin, S. Lukyanov, A. Zamolodchikov and Al. Zamolodchikov,
		\emph{Expectation values of descendent fields in the sine-Gordon model},
		Nucl. Phys. \textbf{B540} (1999) 587–609.
	
\bibitem{Negr_Smir_13-1}
	S. Negro and F. Smirnov,
		\emph{Reflection relations and fermionic basis},
		Lett. Math. Phys. \textbf{103} (2013) 1293-1311, arXiv: 1304.1860.
	
\bibitem{AlZa_06}
	Al. Zamolodchikov,
		\emph{On the thermodynamic Bethe ansatz equation in the sinh-Gordon model},
		J. Phys. \textbf{A39} (2006) 12863-12887.

\bibitem{Lecl_Muss_99}
	A. LeClair and G. Mussardo,
		\emph{Finite Temperature Correlation Functions in Integrable QFT},
		Nucl. Phys. \textbf{B552} (1999) 624-642, arXiv: hep-th/9902075.
		
\bibitem{Luky_01}
	S. Lukyanov,
		\emph{Finite temperature expectation values of local fields in the sinh-Gordon model},
		Nucl. Phys. \textbf{B612} (2001) 391-412, arXiv: hep-th/0005027.

\bibitem{Dorn_Otto_94}
	H. Dorn and H.-J. Otto,
		\emph{Two and three point functions in Liouville Theory},
		Nucl. Phys. \textbf{B429}, 375-388 (1994)

\bibitem{AZam_AlZa_96}
	A. Zamolodchikov and Al. Zamolodchikov,
		\emph{Structure constants and conformal bootstrap in Liouville field theory},
		Nucl. Phys. \textbf{B477} (1996) 577–605, arXiv: hep-th/9506136.
		
\bibitem{AlZa_00}
	Al. Zamolodchikov,
		\emph{On the thermodynamic Bethe ansatz equation in sinh-Gordon model},
		J. Phys. \textbf{A39} (2006) 12863-12887, arXiv: hep-th/0005181.

\bibitem{Bazh_Luky_AZam_96}
	V.V. Bazhanov, S.L. Lukyanov and A.B. Zamolodchikov,
		\emph{Integrable structure of conformal ﬁeld theory, quantum 
KdV theory and thermodynamic Bethe ansatz},
		Commun. Math. Phys. \textbf{177} (1996) 381, 
[arXiv:hep-th/9412229].

\bibitem{Bazh_Luky_AZam_97}
	V.V. Bazhanov, S.L. Lukyanov and A.B. Zamolodchikov,
		\emph{Integrable structure of conformal ﬁeld theory II. 
Q-operator and DDV equation},
		Commun. Math. Phys. \textbf{190} (1997) 247, 
[arXiv:hep-th/9604044]

\bibitem{Bazh_Luky_AZam_99}
	V.V. Bazhanov, S.L. Lukyanov and A.B. Zamolodchikov,
		- \emph{Integrable structure of conformal ﬁeld theory III. The 
Yang-Baxter relation},
		Commun. Math. Phys. \textbf{200} (1999) 297, 
[arXiv:hep-th/9805008].
\end{thebibliography}
\end{document}